\documentclass[twocolumn,showpacs,preprintnumbers,amsmath,amssymb]{revtex4}

\usepackage{graphicx}
\usepackage{dcolumn}
\usepackage{bm}

\begin{document}

\preprint{Submitted to PRC}

\title{Projectile fragmentation reactions and production of nuclei
near the neutron drip-line}%

\author{M.~Notani$^{{\rm 1},}$\footnote{Present address: Physics Division, Argonne National Laboratory, U.S.A.}}
\author{H.~Sakurai$^{{\rm 2},}$\footnote{Present address: RIKEN, Japan}}
\author{N.~Aoi$^{\rm 3}$}
\author{H.~Iwasaki$^{{\rm 2},}$\footnote{Present address: IPN, F-91406 Orsay Cedex, France}}
\author{N.~Fukuda$^{\rm 3}$}
\author{Z.~Liu
$^{{\rm 3},}$\footnote{Present address: University of Surrey, U.K.}}
\author{K.~Yoneda$^{\rm 3}$}
\author{H.~Ogawa$^{\rm 3}$}
\author{T.~Teranishi
$^{{\rm 1},}$\footnote{Present address: Department of Physics, Kyushu University, Japan}}
\author{T.~Nakamura$^{\rm 4}$}
\author{H.~Okuno$^{\rm 3}$}
\author{A.~Yoshida$^{\rm 3}$}
\author{Y.~Watanabe$^{{\rm 3},}$\footnote{Present address: High Energy Accelerator Research Organization (KEK), Japan}}
\author{S.~Momota$^{\rm 5}$}
\author{N.~Inabe$^{\rm 3}$}
\author{T.~Kubo$^{\rm 3}$}
\author{S.~Ito$^{\rm 3}$}
\author{A.~Ozawa$^{\rm 3}$}
\author{T.~Suzuki$^{\rm 3}$}
\author{I.~Tanihata$^{{\rm 3},}$\footnote{Present address: TRIUMF, Canada}}
\author{M.~Ishihara$^{\rm 3}$}
\affiliation{%
$^{\rm 1}$Center for Nuclear Study, University of Tokyo~(CNS) RIKEN Campus, 2-1 Hirosawa, Wako, Saitama 351-0198, Japan}%
\affiliation{%
$^{\rm 2}$Department of Physics, University of Tokyo, 7-3-1 Hongo, Bunkyo, Tokyo 113-0033, Japan}%
\affiliation{%
$^{\rm 3}$RIKEN (The Institute of Physical and Chemical Research), 2-1 Hirosawa, Wako, Saitama 351-0198, Japan}%
\affiliation{%
$^{\rm 4}$Department of Applied Physics, Tokyo Institute of Technology, 2-12-1 Oh-okayama, Meguro-ku, Tokyo 152-8551, Japan}%
\affiliation{%
$^{\rm 5}$Department of Intelligent Mechanical Systems Engineering, Kochi University of Technology, Tosayamada-cho, Kochi 782, Japan}%

\date{\today}

\begin{abstract}
The reaction mechanism of projectile fragmentation
at intermediate energies
has been investigated
observing the target dependence of
the production cross sections of very neutron-rich nuclei.
Measurement of longitudinal momentum distributions
of projectile-like fragments
within a wide range of fragment mass and its charge
was performed using a hundred-MeV/n
$^{40}$Ar beam incident on Be and Ta targets.
By measurement of fragment momentum distribution,
a parabolic mass dependence of momentum peak shift
was observed in the results of both targets,
and
a phenomenon of light-fragment acceleration was found only in
the Be-target data.
The analysis of production cross sections
revealed an obvious enhancement of
the target dependence except target size effect
when the neutron excess is increased.
This result implies the breakdown of factorization~(BOF)
of production cross sections
for very neutron-rich nuclei near the drip line.
\end{abstract}


\maketitle

\section{\label{sec:intro}Introduction}

During these two decades, unstable nuclear physics
has become one of the most interesting fields of nuclear physics.
Nuclear fragmentation of heavy-ion beams is utilized
for producing secondary beams of unstable nuclei far from
$\beta$-\hspace{0pt}stability.
For designing experiments with the secondary beams,
a good knowledge of production cross sections
is essential.
To deduce the production cross sections relevantly,
an empirical parametrization of fragmentation cross sections~(EPAX)
is widely used in simulation programs for projectile-fragment
separators~\cite{Wing92,Iwa97,Sum00}.
For consideration of reaction mechanism,
fragmentation models
based on abrasion-ablation scheme~(AA models)
are often applied to estimate the production cross sections.

With the recent development of heavy-ion accelerators,
the number of accessible nuclei lying far from
$\beta$-\hspace{0pt}stability line has been increasing.
An example has been shown by the experimental findings of
particle stability of
$^{31}$F~\cite{31F}, $^{31,34}$Ne~\cite{Saku96,34Ne},
$^{37}$Na~\cite{34Ne}, $^{37,38}$Mg~\cite{Saku96,Saku97},
$^{40,41}$Al~\cite{Saku97},
and $^{43}$Si~\cite{34Ne},
as the most neutron-rich nuclei so far identified
for the $Z$=9$-$14 elements.
For the new isotope-search experiment,
the accuracy to predict the production cross section of
neutron-rich nuclei near the drip line
is very important to discuss the particle stability of them.
However, the predictive power of the EPAX parametrization and AA models
is not strong enough
for specific very neutron-rich nuclei~\cite{Gaim91}.
For instance,
although tantalum is often experimentially used as a production target
in order to earn the better yield of these nuclei,
the target dependence of production cross sections
is not taken into account in these models.
Instead,
we first determined the production cross sections for the observed isotopes,
which were then used to estimate the production cross sections
and the expected yields for the non-observed isotopes.
Therefore,
to enhance predictive powers for production cross sections,
deeply understanding reaction mechanism as well as observing
systematic behaviors of production cross sections are necessary.

The validity of the EPAX parametrization and AA models
has been mainly verified for medium and heavy mass fragments
via multi-GeV high-energy fragmentation reactions.
The EPAX formula has been also used for the
intermediate-energy experiments~(several tens $A$ MeV)
since the formula can reproduce reasonably well even
at intermediate enegies
the production cross sections
of stable and unstable nuclei near the $\beta$-stability line.
A target dependence of fragment-production cross sections
in relativistic heavy-ion collisions is limited only
by taking account of the nuclear-size effect~\cite{Cum90}.
However, the recent experiments show that
the production yields of nuclei far from stability line
are quite different from the prediction of EPAX formula,
and strongly dependent on the $N$/$Z$ ratio of
the target~\cite{Guer90}.
It is of great interest
whether the cross sections measured with different targets factorize
in projectile fragmentation at intermediate energies.

In order to confirm the validity of factorization,
a careful measurement of fragment momentum distribution
is necessary for precise determination
of the production cross sections.
At relativistic energies, the shape of momentum distribution
for an isotope was found to be a Gaussian function~\cite{Heck72}
and the width was well understood
with a statistical model~\cite{Gold74}.
On the other hand, the momentum distribution of fragments
produced at intermediate energies
has an asymmetric shape
with a tail at low momentum side.
A theoretical attempt
was made to reproduce the asymmetric shape,
by taking into account nucleon flows
between projectile and target in the collision time~\cite{Got92}.
In this model,
stochastic nucleon transfers using a Monte Carlo method
and sequential evaporation were taken into account.
This model can reproduce the low-momentum-side
tails, however, due to a large friction force, the whole of
predicted distributions are shifted toward the low momentum side
much larger than the observed.
This discrepancy of momentum distributions
affords a large ambiguity
to evaluate the production cross section
from the measured yield of fragments.
In addition,
the measurement for very neutron-rich nuclei
has been performed using very thick targets to earn the yields, so far.
As distortions of momentum distributions due to target thickness,
the momentum distributions of fragments cannot be obtained from
these data clearly.
Therefore, the measurement of momentum distribution of
very neutron-rich nuclei has become important.

In the present work,
we focus on the target dependence of momentum distribution
of projectile-like fragment~(PLF) produced by nuclear fragmentation
reactions at an intermediate energy.
To investigate the target dependence of the production cross sections
systematically,
we used two production targets of Be and Ta.
To avoid distortions of momentum distributions due to the
target thickness, we prepared relatively thin targets.
We performed the experiment with the RIKEN-RIPS
to eliminate the primary beam and to collect the projectile-like fragments.
The data of this experiment were taken
in a wide range for fragment mass including
very neutron-rich side~(${N/Z}_{f}$ $\approx$ 3) toward the neutron drip-line
and light mass~(${A}_{f}$ $\geq$ 3),
with
a good statistics for momentum-distribution tails.

In the following, we first describe the experimental setup and
procedure in Sec.~II. In Sec.~III, the analysis of the data
is described.
In Sec.~IV, the experimental results are
described, where the observed momentum distributions of fragments,
the momentum peak shift, the high- and low-momentum side widths,
and the production cross sections
are presented in terms of target dependence.
Based on the results, we discuss 
the prefragment production mechanism
in projectile fragmentation reactions.

\section{\label{sec:setup}EXPERIMENTAL SETUP AND PROCEDURE}

The projectile fragmentation experiment using a $^{40}$Ar beam
was performed at the RIKEN Accelerator Research Facility.
The measurement of momentum distributions
of projectile-like fragments was carried out
with the projectile fragment separator, RIPS~\cite{Kubo92}.
The $^{40}$Ar${}^{17+}$ beam accelerated by the ring
cyclotron with energies up to 90\/$A$ MeV and 94\/$A$ MeV
irradiated
a 95-mg/${\rm {cm}^{2}}$ thick $^{9}$Be target and
a 17-mg/${\rm {cm}^{2}}$ thick $^{nat}$Ta target,
respectively.

The primary beam intensity was monitored for normalization
of fragment yields to obtain the momentum distribution.
A plastic telescope consisted of three plastic scintillators
with a size of 50$\times$50$\times$0.5 mm$^{3}$
was placed at
a backward angle of 135 degrees and
at a distance of 0.5 m from the target position.
The plastic detectors counted yield rates of light particles
produced with nuclear reactions at the production target.
The three photomultiplier tubes~(PMTs) were mounted on
the plastic scintillators one-by-one.
The counting rate of triple coincidence was used to
monitor the primary beam.
The primary beam intensity was measured with
the indirect method of plastic counters calibrated by
a Faraday cup.
By changing the primary-beam intensity
which ranges from ${10}^{-4}$ to 1.0 of the full beam intensity,
the calibration data were taken.
The systematic error of the beam monitor was estimated
to be 7\%.

 The RIPS was used as a doubly achromatic spectrometer.
Projectile fragments produced at the production target and
emitted at 0\raisebox{1ex}{$\circ$} were collected
and transported to a double achromatic focal plane~(F2).
The momentum acceptance was set to be ${\Delta}p/p$=1\%
at a momentum dispersive focal plane~(F1)
where left and right slits formed the rectangle window of
momentum acceptance.
The angular acceptance was set with a square window
formed by four slits (upper, lower, left and right), which were placed
at the behind of the production target.
The $\theta$ and $\phi$ angular acceptances were
25 mrad,
which is narrower than the width of angular distribution of
fragments in r.m.s.
We use the constant value of 6.25$\times {10}^{-1}$ msr
as the solid angle of $\Delta\Omega$.

Momentum distributions of fragments were measured
at 23 settings of magnetic rigidity~($B\rho$)
over a range of 2.520$-$4.068 Tm using $^{9}$Be target.
In the case of $^{181}$Ta target, the measurement of
momentum distribution for each fragment was performed
at 31 magnet settings as same as Be case.
When the magnetic field was changed for each run,
the F2 image of secondary beam was measured by means of
a parallel-plate avalanche counter (PPAC)~\cite{Kuma01}
to confirm transmission,
and the x-position of the beam at F2 was corrected to center
precisely by tuning the D2 field.
The beam position was monitored with an accuracy of 1 mm.
The systematic error of magnetic rigidity was
about 3$\times {10}^{-4}$ from the ratio of 1 mm to 3.6 m.
The difference between D1 and D2 magnetic fields was less than 0.05\%.

The identification of fragments was carried out
event-by-event by means of measurement of
time of flight (TOF) and energy deposit ($\Delta E$)
for each fixed $B\rho$ run with the 1\% momentum slit.
According to an estimation of charge state distribution~\cite{Bad91},
all fragments in flight are fully striped ($Q$\/$\cong$\/$Z$).
Under this assumption,
the particle identification can be performed on the basis of the relations:
\begin{eqnarray}
B\rho & \propto & \frac{A}{Z}\beta\\
\Delta E & \propto & {(\frac{Z}{\beta})}^{2}\\
\beta & \propto & \frac{1}{TOF},
\end{eqnarray}
\noindent
where $A$ and $Z$ are mass number and atomic number,
respectively.

The detectors of two 0.5-mm thick surface-barrier-type
silicon counters (SSD1, SSD2) and
a 0.5-mm thick plastic scintillation counter (PL) were installed
for the $\Delta E$ and TOF measurement
at F2.
To use two silicon detectors allowed us to deduce $Z$ number
of the fragment
independently, and to take correlations between them
to achieve good $S$/$N$ ratio.

Both SSD1 and SSD2 have 48 mm$\times$48 mm sensitive area
which is wide enough to accept all particles reaching to F2
where FWHM of the beam profile is 6 mm.
Two photomultiplier tubes~(PMTs) were mounted on both sides of
PL (Left and Right). A timing of PL was determined with
an average of the signals from the Left and Right.

The TOF of each fragment over 21.3-m flight path between
the production target and the F2
was determined from the difference of timing signals
between RF signal of the cyclotron and the PL timing.
The TOF resolution was measured with a faint beam
to be 0.27 ns (r.m.s.), which included
the timing jitter of RF signal~($\sim$0.09 ns).
Thus, the intrinsic resolution for PMT was estimated to be 0.18 ns.

\section{\label{sec:analys}DATA ANALYSIS}

In this section we describe the procedures
of the data analysis, fitting of momentum distributions,
and evaluation of the production cross sections.

\subsection{\label{sec:pi_procedure}Particle identification}

\begin{figure}[t]
\scalebox{.4}{\includegraphics{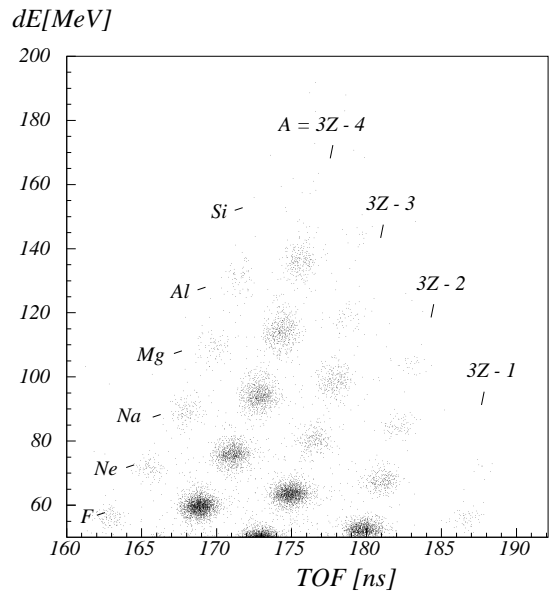}} 
\caption{\label{fig:dETOF}~Particle identification in the dE-TOF plane
at $B\rho$ = 3.708 Tm with the Be target.}
\end{figure}

\begin{figure}[h]
\scalebox{.45}{\includegraphics{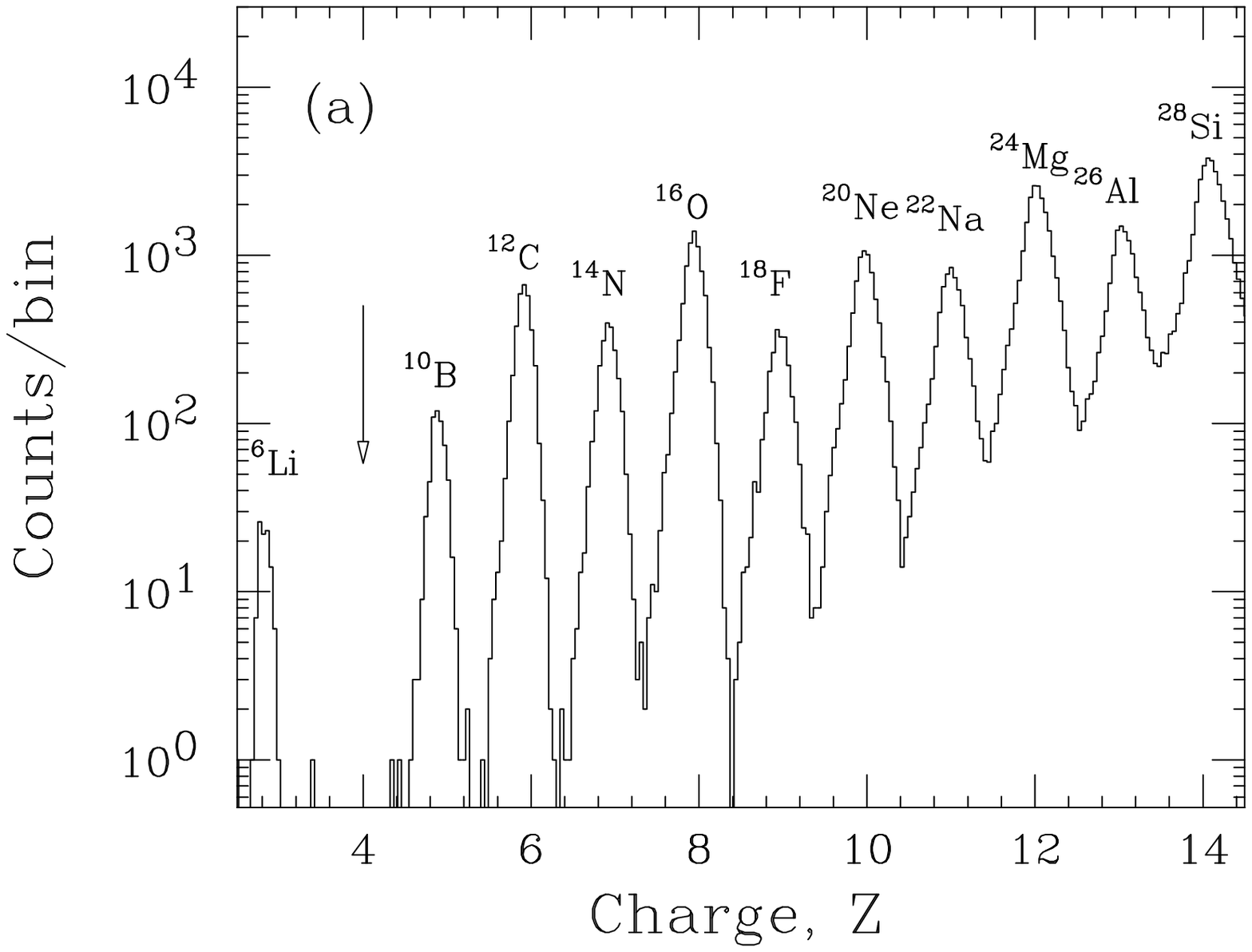}} 
~\vspace{7pt}\\
\scalebox{.45}{\includegraphics{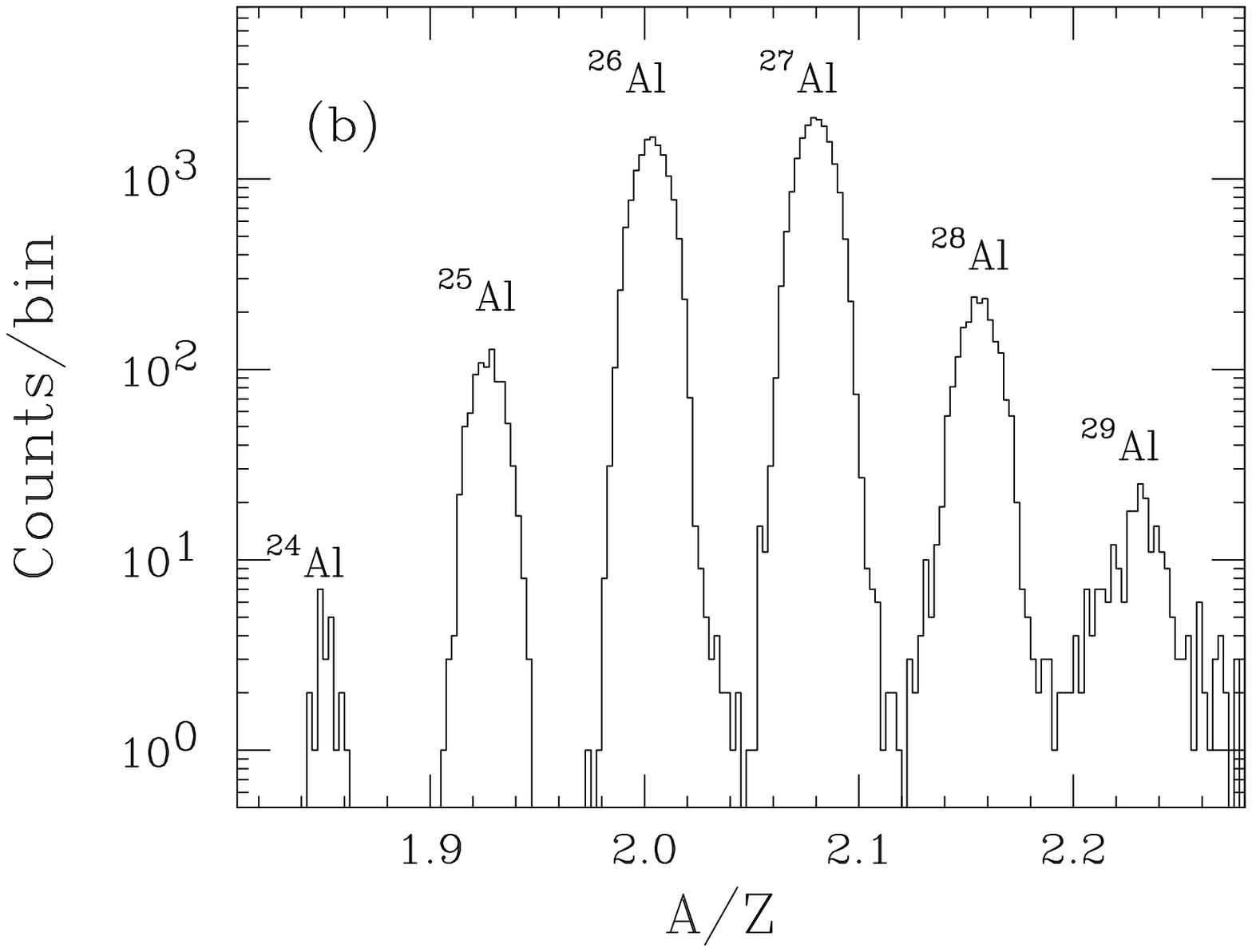}} 
\caption{\label{fig:run31-proj}~(a)~$Z$-projection spectrum
for $A$/$Z$=2$\pm$0.3 and
(b)~$A$/$Z$-projection spectrum for $Z$=13.0$\pm$0.5 
at $B\rho$=2.523 Tm (Be target). The arrow in (a) indicates the lack of
$^{8}$Be, which is known to be particle unbound.}
\end{figure}

Figure~\ref{fig:dETOF} shows
a two-dimensional plot versus $\Delta E$ in SSD1
for one $B\rho$ setting using the Be target.
A rejection of the background events
was carried using a correlation gate between SSD1 and SSD2.
By use of the 3-$\sigma$ gate by two SSDs,
we achieved the particle identification of fragments
with low background events.
Fragment yields were obtained by counting the isotopes
from the particle identification.
Figure~\ref{fig:run31-proj} shows
the $Z$- and $A$/$Z$-projection of the particle identification
at $B\rho$=2.523 Tm setting using the Be target.
The counting gate of isotopes was a rectangle region with
$\pm$3-$\sigma$ of the resolution~$\sigma$(r.m.s.) for $Z$ and $A$/$Z$.

In the data acquired using the Be target, the analyzed isotopes were
$^{6-9}$Li,
$^{7-12}$Be,
$^{10-15}$B,
$^{11-18}$C,
$^{13-21}$N,
$^{15-24}$O,
$^{17-27}$F,
$^{19-29}$Ne,
$^{21-32}$Na,
$^{23-34}$Mg,
$^{25-36}$Al,
$^{27-38}$Si,
$^{29-39}$P,
$^{33-38}$S,
$^{36-39}$Cl, and
$^{39}$Ar.
Most of these isotopes are neutron-rich ones.
It should be noted that the 
$^{36}$Al,$^{37,38}$Si, and $^{38,39}$P isotopes analyzed have
neutron numbers larger than the projectile (N $\geq$ 23),
which are produced through neutron pick-up process.
We analyzed from the Ta-target data for
$^{6-8}$Li,
$^{9-11}$Be,
$^{10-14}$B,
$^{11-17}$C,
$^{13-19}$N,
$^{15-21}$O,
$^{17-24}$F,
$^{19-27}$Ne,
$^{21-29}$Na,
$^{23-31}$Mg,
$^{24-34}$Al,
$^{26-34}$Si,
$^{29-36}$P,
$^{30-37}$S,
$^{33-37}$Cl,
$^{35-39}$Ar, and
$^{37-40}$K.
The potassium isotopes should be produced by the reaction with
proton pick-up process.

In order to obtain the doubly-differential cross section
from each fragment yield,
we estimated the transmission between F0 to F2,
and the reaction loss in the detectors.
A Monte Carlo simulation by MOCADI~\cite{Iwa97}
was performed under the realistic condition of RIPS
using a reference beam of $^{40}$Ar.
The value of transmission was obtained to be 95.3$\pm$0.3\%.
The nuclear reaction loss of fragment in the detectors
was evaluated with reaction cross sections
calculated by a simple geometrical model.
The reaction loss in the detectors was estimated to be less than 0.8\%.
After all, the systematic error was $\pm$9\%
for evaluation of the fragment cross sections.

\subsection{\label{sec:fit_procedure}Fitting Procedure}

Momentum distributions of fragments have information
for understanding reaction mechanisms.
At relativistic energies, the projectile fragments have symmetric
momentum distributions fitted to a Gaussian form,
of which width has been discussed with respect to the Fermi
motion of nucleons or temperatures of pre-fragments~\cite{Gold74}.

\begin{figure}[t]
\scalebox{.45}{\includegraphics{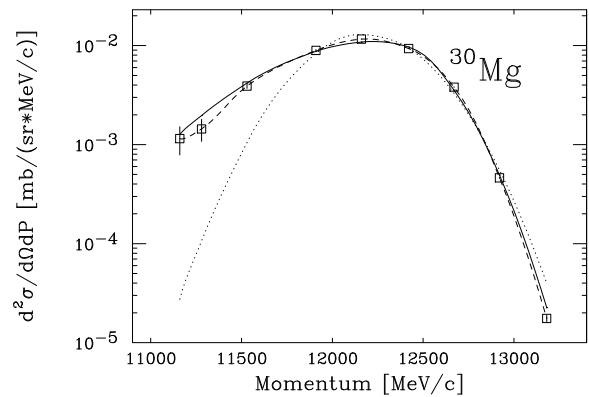}} 
\caption{\label{fig:mg30}
Typical fragment momentum distribution (dashed curve)
and fitting results for the momentum distribution
of $^{30}$Mg data.
The fitting result with a Gaussian function~(dotted curve)
shows clearly asymmetric feature of the experimental data.
The solid curve indicates a fitting result
with the asymmetric Gaussian function.}
\end{figure}

The results obtained in this work are different from
at relativistic energies. Figure~\ref{fig:mg30} shows
a typical momentum distribution of this experiment.
By comparison of the momentum distribution fitted
with a Gaussian function~(dotted curve),
asymmetric feature of the distribution is clearly observed.
The momentum distribution of projectile-like fragments
produced at intermediate energies
are generally asymmetric with a tail on the low momentum side~\cite{PFA95}.

To deduce the most probable momentum and width from such skewed shapes,
the momentum distributions have been fitted so far
with several kinds of trial functions~\cite{PFA95,Baz90,Bac93}.
Since physical models have not been established for
the low momentum tail, what kind of functions to be used
is not unique.
To study a systematics of low-momentum tail,
the following asymmetric function with four free parameters
is applied for the present data fitting,
\begin{eqnarray}
\frac{{d}^{2}\sigma}{dPd\Omega} (\theta=0\mbox{\raisebox{1ex}{$\circ$}}) =
\left\{%
\begin{array}{c}
\displaystyle
A \cdot \exp(-\frac{(P-P_0)^2}{2{\sigma}_{L}^{2}}), P\le P_0\\
\displaystyle
A \cdot \exp(-\frac{(P-P_0)^2}{2{\sigma}_{H}^{2}}), P\ge P_0
\end{array}\right.\;,
%
\label{eq:one}
\end{eqnarray}
where 
${P}_{0}$ is the most probable peak value of momentum
in the distribution,
${\sigma}_{L}$ and ${\sigma}_{H}$
are momentum width in low and high energy side,
respectively.

 In this fitting procedure, the maximum likelihood method was
used to treat the small statistics at the tail parts of distribution.
The result of fitting with the asymmetric Gaussian function
is shown with the solid curve in Fig.~\ref{fig:mg30}.

\begin{figure}[t]
\scalebox{.55}{\includegraphics{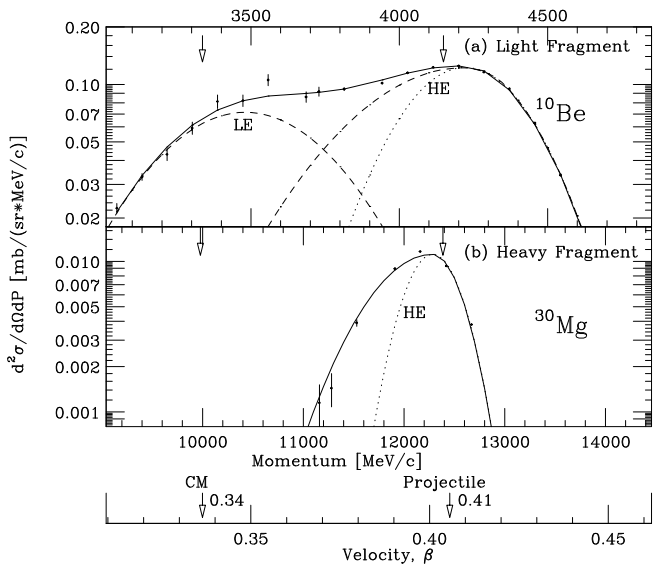}} 
\caption{\label{fig:typical}~Typical fragment momentum distributions
for (a)~$^{10}$Be and (b)~$^{30}$Mg produced in Ar+Be reaction.
The fitting results are also shown with solid curves.
Light fragment of $^{10}$Be has two components~(dashes curves)
of HE- and LE-components, while heavy fragment of $^{30}$Mg has
one component corresponding to HE-component.
The HE-component for each isotope has a low momentum tail
and the symmetric parts are shown with dotted curves.}
\end{figure}

By means of this method,
we obtained fitting results of momentum distribution for all isotopes
available in our data,
though there is another kind of complexity in a few case.
We found two components in the momentum distributions of
very light fragments only for the Be-target data.
Figure~\ref{fig:typical} shows the momentum distributions
of the $^{10}$Be and $^{30}$Mg isotopes from
the Be-target data.
The both distributions are
scaled as a function of velocity~($\beta$).
Two arrows in the figure indicate the velocities
of projectile~(${\beta}_{proj}$ )
and center-of-mass system~(${\beta}_{cm}$), respectively.
In the distribution of $^{30}$Mg~(Fig.~\ref{fig:typical}(b)),
a single component is observed
near the projectile velocity~(${\beta}_{proj}$).
On the other hand, the $^{10}$Be distribution~(Fig.~\ref{fig:typical}(a))
shows two components at ${\beta}_{proj}$ and ${\beta}_{cm}$.
Here, the component around ${\beta}_{proj}$ is defined as
a high-energy side of peak~(HE), and that of ${\beta}_{cm}$
is as a low-energy side of peak~(LE).
We made an attempt to fit the data
using the asymmetric Gaussian function
for the HE-component and a Gaussian function for the LE-component.
The fitting results are drawn with solid curves.
With increasing the fragment mass number, the LE-component decreases
very quickly.
At last, no significant LE-component has been observed for
heavy fragments like the $^{30}$Mg data.
In our data, we have found clearly the LE-component
for the fragments with $A$ of 9$\sim$12.
The LE-component has been observed for light fragments
in Ar+Be reaction,
while,
the LE-component has not been found in the momentum distributions
of Ar+Ta reaction data in the momentum region
where the experimental data were taken.
When a light fragment like $^{10}$B is produced in
the Ar+Ta reaction, the impact parameter is much larger
than in the case of the Ar+Be reaction.
Thus, we found the LE-component only for the Ar+Be system.
Since we focus on the projectile fragmentation reaction,
we discuss mainly the HE-component in this paper.
%
%

\subsection{\label{sec:evaluation_cs}Evaluation of cross sections}

The production cross sections of fragments were evaluated
with the fitting results
of the longitudinal momentum distributions.
The transverse momentum distributions are assumed
to be a Gaussian function with a width of ${\sigma}_{\perp}$,
\begin{eqnarray}
{\sigma}_{\perp}^{2} & = &
{\sigma}_{H}^{2} +
\frac{{A}_{f}({A}_{f}-1)}{{A}_{p}({A}_{p}-1)}{\sigma}_{D}^{2},
\label{eq:transverse}
\end{eqnarray}
\noindent
where ${\sigma}_{D}$ is a parameter of the deflection effect,
and ${A}_{p,f}$ are mass numbers of a projectile and a fragment,
respectively. 
%
\begin{figure}[t]
\scalebox{.45}{\includegraphics{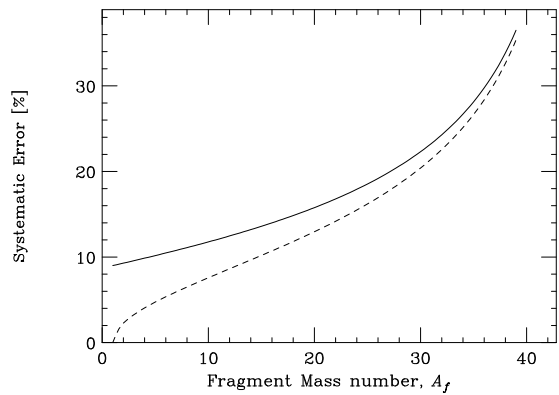}} 
\caption{
\label{fig:cs-syserror2}~Overall systematic error
for the production cross sections~(solid line).
The dashed line shows error originated from
ambiguity of ${\sigma}_{\perp}$.}
\end{figure}
In Ref.~\cite{Van79}, the deflection parameters for two targets
of $^{27}$Al and $^{197}$Au with 92.5 and 117.5$A$ MeV $^{16}$O beams.
The results from their measurement were
${\sigma}_{D}$=190\/$-$\/220 MeV/c, found no large target dependence.
Taking account of an energy dependence reported in Ref.~\cite{Day86},
we used ${\sigma}_{D}$
of 195 MeV/c around 90$A$ MeV for the present experiment.
The ambiguity of transverse momentum distributions
was taken into account as the systematic error of
the cross sections.
The overall systematic error for the production cross sections
is shown in Fig.~\ref{fig:cs-syserror2}.

\section{\label{sec:result}Results and Discussion}

The fitting results with the asymmetric Gaussian function 
are presented in the following.
First of all, the fitting results of momentum distributions
are compared with several formulae taken from reaction models.
The result of high momentum
side width can be understood by the Goldhaber model.
However, we need the further discussion of the momentum peak shift
and low momentum side width.
Next, we show the result of production cross sections. The result
reveals the phenomena, breakdown of factorization~(BOF),
for production of very neutron-rich nuclei.
We discuss the systematics of isotope production cross sections.
The charge distribution of the cross sections for a fragment mass
is obtained from our data,
and compared with the EPAX formula.
Finally, we discuss the prefragment production mechanism in
projectile fragmentation reactions
to search for the origin of BOF.

\subsection{\label{sec:peak_shift}Momentum peak shift}

In the nuclear fragmentation process,
a part of the kinetic energy of projectile is converted into
excitation energies of fragments, and the projectile velocity
is decreased. 
This energy loss in projectile fragmentation reaction
is called as `momentum peak shift'.
The momentum peak shift is obtained from
the difference of the projectile velocity and the most probable
velocities of fragments.
We present the result of momentum peak shift in a unit of energy
per nucleon, which is proportional to square of velocity.
This unit is convenient to discuss the kinetic energy consumption
in the nuclear reaction.

\begin{figure}[t]
\noindent
\scalebox{.45}{\includegraphics{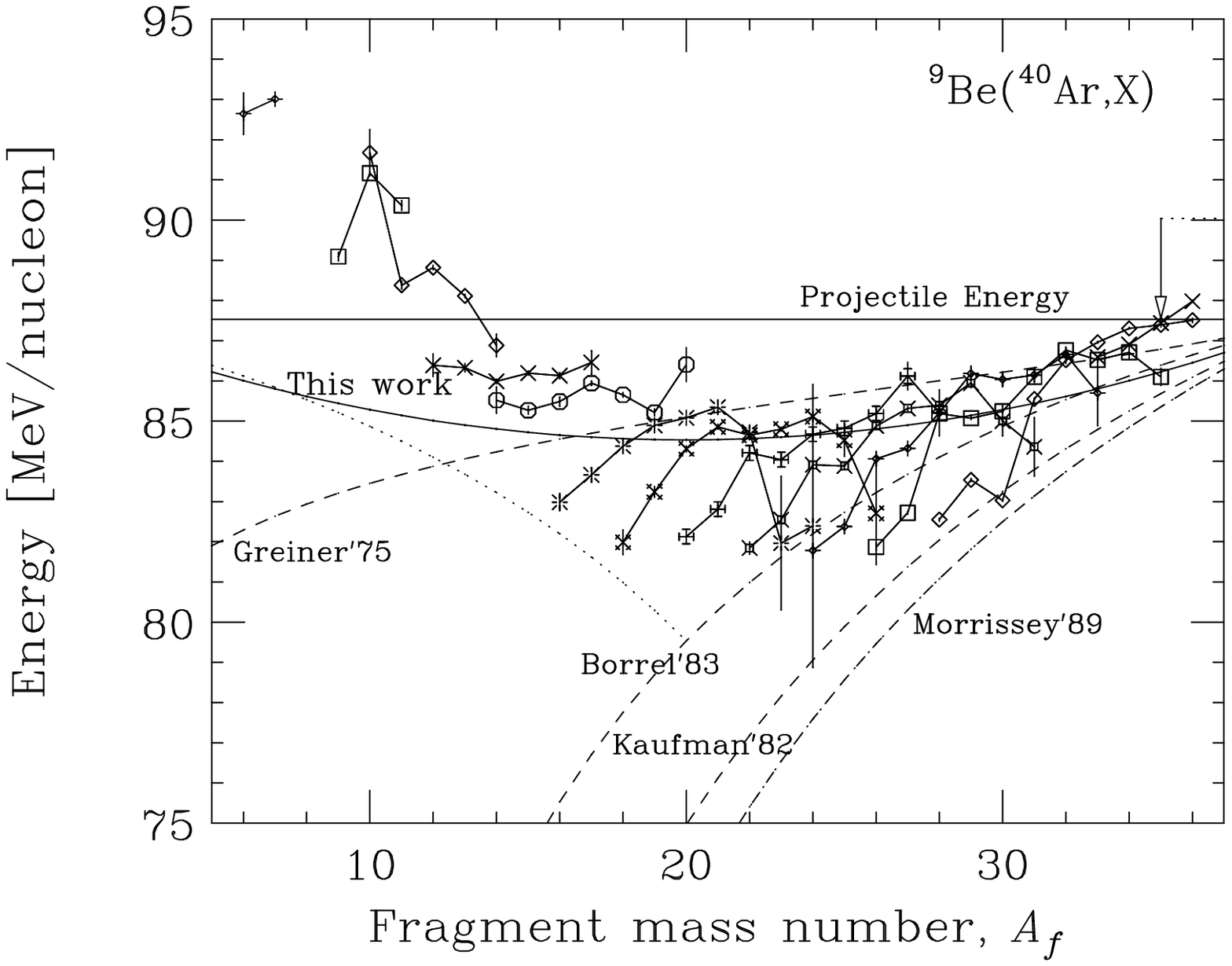}} 
~\vspace{7pt}\\
\scalebox{.45}{\includegraphics{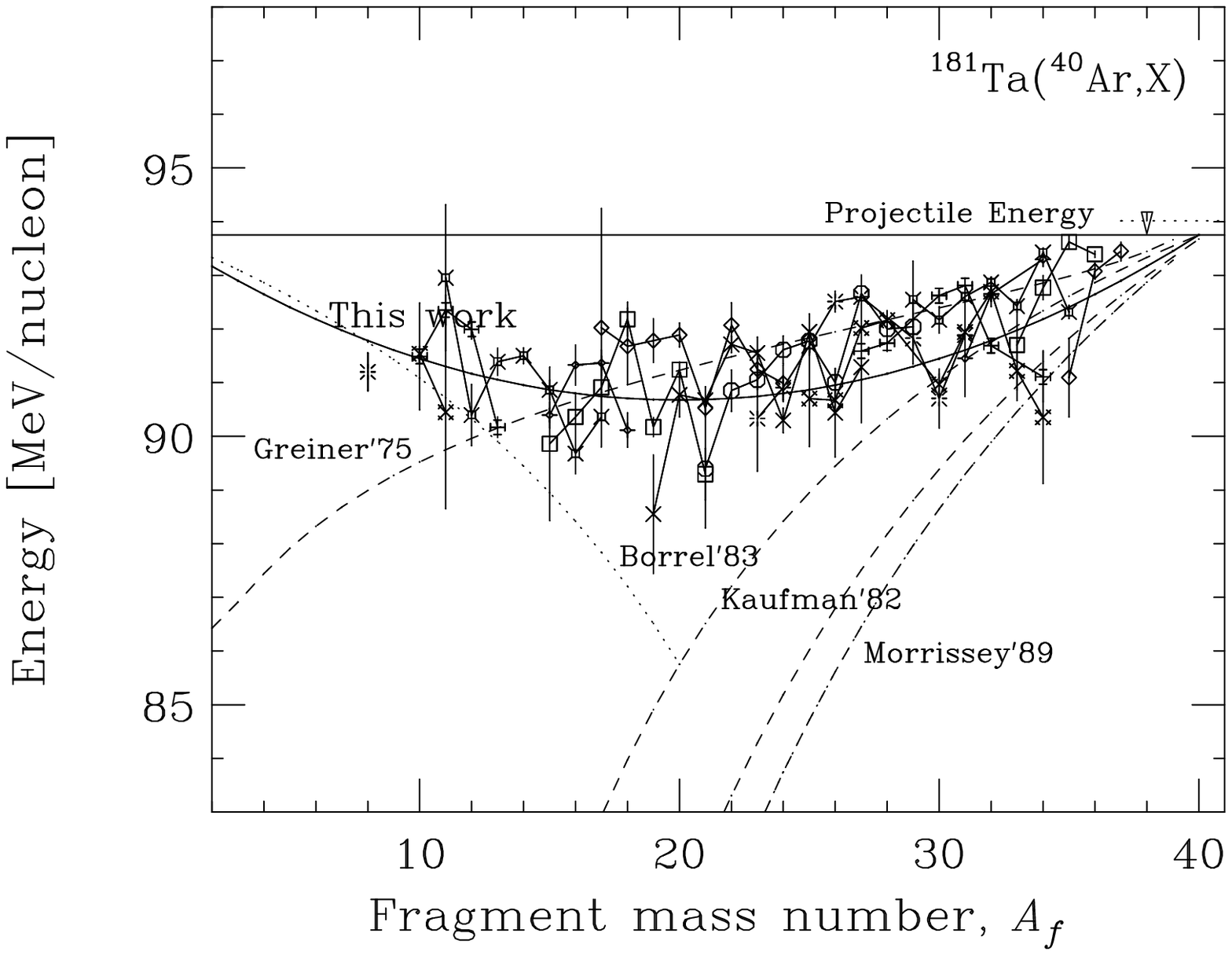}} 
\caption{\label{fig:mps}~Momentum peak shift of fragments produced
in (upper) Ar+Be and (lower) Ar+Ta.
The primary beam energies are drawn with dotted lines, and
the kinetic energies at the half point of
target thickness~(projectile energies) are the solid lines.
See the text for the labels and curves.}
\end{figure}

Figure~\ref{fig:mps} shows the momentum peak shift of fragments
produced in Ar+Be and Ar+Ta reactions.
The solid lines for each isotope are drawn to guide the eye.
The primary beam energies were corrected with the mean energy loss
in the production target.
The obtained projectile energies~(solid lines) were taken to be
87.5$A$ MeV and 93.8$A$ MeV, respectively.
The values of the most probable energies of fragments were also
corrected with energy losses in the targets.
The negative shift of data from the solid line of the projectile
energy indicates the energy loss by nuclear fragmentation process.
%
The systematic error of momentum shift was estimated to be 0.9\%
in the unit of energy per nucleon.

The deviation for a given element chain in the Be target
is comparable to the systematic error.
As seen in Fig.~\ref{fig:mps}, the measured momentum shifts
for the Be target~(a), compared with the case of the Ta target~(b),
vary widely.
One could understand the deviation as an effect of the target thickness
in atomic energy-loss process~\cite{Dufo86},
because the target thickness of Be is six times thicker than that of Ta.

The bounce of momentum peak shift as a function of fragment mass
was observed in the results of both targets.
The momentum peak shift increases when the number of removed nucleons
$\Delta A={A}_{p}-{A}_{f}$ is increased up to a half of projectile mass
(${A}_{f}\geq$20).
On the other hand,
the momentum shift decreases
when the mass loss $\Delta A$ is increased beyond 20 (${A}_{f}<$20).
In short, we observed the maximum of momentum peak shift
around ${A}_{f}$=20.

A phenomenon of fragment acceleration was observed in the Be-target data.
In Fig.~\ref{fig:mps}\/(a), the solid line of 87.5$A$ MeV
corresponds to the primary beam energy.
Velocities of the fragments $^{6,7}$Li, $^{9-11}$Be, and $^{10-13}$B
are larger than the projectile velocity.
The very light fragments are accelerated
in the reaction process.
We note that the acceleration phenomena was 
also observed 
in collisions of $^{238}$U at 1 A GeV with lead,
reported by Enqvist~\cite{Enqvist99}.
On the other hand, no acceleration phenomenon for all fragments
${A}_{f}\geq$8 was observed in the Ta-target data.

Except the acceleration phenomenon,
no significant difference between both targets
was observed in the momentum peak shift
for the projectile-like fragments ${A}_{f}\geq$20.

We found two features in the momentum peak shift.
First, the maximum of momentum peak shift is observed for the fragments
around ${A}_{f}=$20 in Fig.~\ref{fig:mps} for both targets.
Secondly, the acceleration phenomenon was observed
in light fragments only for the Be target.
Namely, the most probable velocities of light fragments are
beyond the projectile velocity.
In the following, we discuss the features observed in the
momentum peak shift.

\subsubsection{\label{sec:parabolla}Parabolic mass dependence of peak shift}

The momentum peak shift has been investigated for long time.
Many of the reports have shown a linear mass dependence of
the peak shift for fragments with ${A}_{f}\geq {A}_{p}/2$.
Several formulae
are proposed so far to reproduce the momentum peak shift
for the heavy fragments produced in peripheral collisions.
We compare our experimental results for the wide fragment mass range
with the formulae and make an attempt to
introduce a new picture to reproduce
the parabolic mass dependence.

In Fig.~\ref{fig:mps},
four formulae are shown with dashed curves and
the labels as Greiner75, Kaufman82,
Borrel85 and Morrissey89~\cite{Grei75,Cum80,Cum81,Kauf82,Morr89,Bor83,Bor86}.
For every formula,
the momentum shift becomes large when the number of removed nucleons
$\Delta$$A$ is increased up.
This tendency conflicts the present result in the region of
${A}_{f}\leq$20.
To compensate for the deviation,
J.A.Winger et al.~\cite{Wing92}
obtained a formula symmetrized
mathematically with respect to ${A}_{p}$/2~(dotted line).
However, it still overestimates the momentum shift
around ${A}_{f}\sim{A}_{p}/2$.
After all, all of formulae above
cannot predict the parabolic mass dependence.

The previous investigations were mainly performed
in the region of fragments close to the projectile mass.
For whole fragment mass region observed in this work,
a parabolic dependence,
where the symmetric point is at the half mass of projectile,
is observed.
As Borrel commented first in Ref.~\cite{Bor86},
the symmetric behavior
of the velocity shift implies another mechanism less costly,
compared with the removal of individual nucleons.
The mechanism may be a process that the projectile splits into two pieces.
Then we make a new formula which calculates the momentum shift
based on the splitting picture.

First, we assumed that the projectile splits in two fragments
which are not exited state.
By means of
the empirical mass formula of Weizs\"acker-Bethe
$M$($N$,$Z$),
where ($N$,$Z$) is the neutron and charge number of nuclei~\cite{BM58},
the energy loss in the splitting process is described as
$M({N}_{f},{Z}_{f})+$
$M({N}_{p}-{N}_{f},{Z}_{p}-{Z}_{f})-$
$M({N}_{p},{Z}_{p})$.
The change of binding energy shows the parabolic mass dependence
of peak shift.
However, the energy loss by the splitting process gives
only several tens percent of the value
measured as momentum peak shift, quantitatively.

Next,
we deduce a semi-empirical formula
in consideration of the excitation energy of fragments.
The splitting process consumes
the kinetic energy of projectile due to the excitation.
Here we assume that the energy consumption
is proportional to the number of pairs of nucleons
destroyed in the reaction,
where the nucleons are acting mutually by long range force.
In this picture, the number of pairs of nucleons in projectile
is counted with
$_{Ap}C_{2}$=${A}_{p}$(${A}_{p}$$-$ 1)/2,
where the mark of $_{n}C_{m}$ is the combinatorial
which gives the number of ways of choosing m out of n.
When the projectile splits into the spectator ${A}_{f}$
and the participant~(${A}_{p}-{A}_{f}$),
the number of pairs of nucleons which decreased by reaction
serves as
${A}_{f}$(${A}_{p} - {A}_{f}$).
When one nucleon is removed~(${A}_{f}$=${A}_{p}-$1),
the number of pairs which decreased
is ${A}_{p}-$1,
and the total energy loss can be defined as
$\varepsilon$~MeV.
The energy loss per a nucleon pair
is $\varepsilon$/(${A}_{p}-$1)~MeV on average.
Thus, the kinetic-energy loss in the splitting process can be written as,
\begin{eqnarray}
\Delta E & = & \frac{\varepsilon {A}_{f}({A}_{p}-{A}_{f})}{{A}_{p}-1}.
\label{eq:eloss-formula}
\end{eqnarray}
\noindent
In the case of non-relativistic beam energy,
the energy conservation between projectile and two pieces gives
a new formula as follows:
\begin{eqnarray}
\frac{{v}_{f}}{{v}_{p}} & = & \sqrt{1~-~\frac{{\varepsilon}~{{A}_{f}}({A}_{p}-{A}_{f})}{{A}_{p}{E}_{p}({A}_{p}-1)}}.
\label{eq:notani-formula}
\end{eqnarray}
\noindent
If we select the energy loss parameter of $\varepsilon$=8 MeV,
this formula corresponds to Borrel's for the case of one-nucleon removal.
However, the value of $\varepsilon$ may not be 8 MeV.
In Fig.~\ref{fig:mps}, the ${P}_{0}$ values predicted by the formula
were compared with the present data.
The solid curve~(this work) was drawn with $\varepsilon$=12 MeV
to fit the data.
The experimental data may support that
$\varepsilon$=12 MeV is better than 8 MeV.

\subsubsection{\label{sec:accel}Acceleration phenomenon}

The acceleration effect cannot be explained by
a fragmentation model based on the abrasion-ablation picture,
in which the projectile always loses the kinetic energy
in the laboratory system for abrasion process of nucleons.

The acceleration phenomenon may be peculiar to
the $^{9}$Be($^{40}$Ar, X) reaction, namely inverse kinematics.
No acceleration effect was found in the Ta target data.
By comparison between both targets,
we found two particular features of momentum distributions
for light fragments produced in the Ar+Be reaction.
One feature is the acceleration phenomenon.
The other feature is the existence of LE-component.
Both phenomena coincide in our data.
Therefore,
the acceleration phenomenon may be related to the LE-component.

F.~Auger et al.~\cite{Aug87} have suggested that the light fragments
originate in division of a composite system.
In their experiment of $^{86}$Kr+C and $^{86}$Kr+Al collisions,
the energy of the Kr beam was 35$A$ MeV,
and relatively lower than our experiment.
As a result of their experiment,
two components of fission-like fragments
were found in the velocity distribution
of fragments with each mass number.
The two components may be generated from
a highly excited compound system
since the average speed of them was equal to
the center-of-mass~(CM) speed of the incident system.
In short, it is possible that the LE-component and
the HE-component come from one source.

%
%
\begin{figure}[t]
\scalebox{.45}{\includegraphics{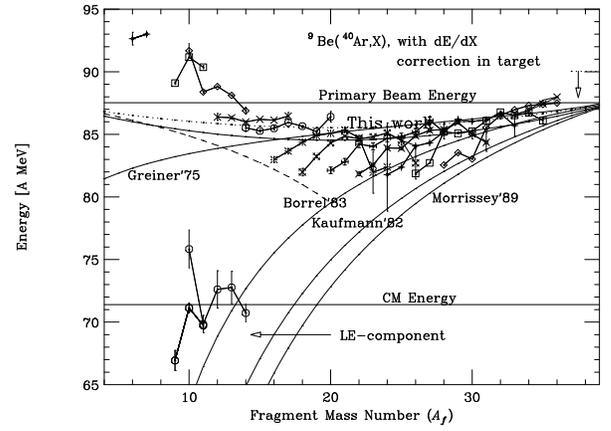}} 
\caption{\label{fig:Accel}~Peak value, ${P}_{0}$, of fragment
momentum distribution for HE-component and LE-component.}
\end{figure}
%
As shown in Fig.~\ref{fig:Accel}, the velocities of
the LE-component observed in this work
are not below but just on the CM velocity.
In addition, the mean energy of HE- and LE- component is
around 80$A$ MeV.
Thus, the observed phenomenon is different from
the work of F.~Auger et al.

Then, we suggest two source model that
the LE-component and the accelerated HE-component
come from two sources produced in different reactions, respectively.
One source is an excited compound system
from projectile and target nuclei running on the CM velocity.
The source emits light particles homogeneously in angular space
and becomes a fragment in the LE-component.
On the other hand, the other source is a hot projectile.
The hot projectile is formed by a strong impact of target nucleus
to convert the kinetic energy of projectile nucleus
to the internal energy for the excitation of projectile nucleus.
The hot projectile decays to two components again,
which consists of a faster component and a slower component
than the projectile velocity,
like the one-source model.
This two-source model predicts a third component
in the momentum distribution of light fragments.
Such third component was not observed in our data.
After all, the origin of the acceleration phenomenon
has not been understood so far.


Neither the acceleration effect nor the LE-component
were found in the Ta target data.
The different situation from the Be target
comes from the large impact parameter.
For a fragment mass,
the impact parameter in the Ar+Ta reaction is larger than in the Ar+Be reaction
when the fragment is formed by the geometrical cut.
The hot compound system is not produced
in the peripheral collision with the large impact parameter.
Therefore,
the exotic phenomena may not be observed in the Ta target data.

\subsection{\label{sec:h_width}High momentum side width}
%
\begin{figure}[t]
\scalebox{.40}{\includegraphics{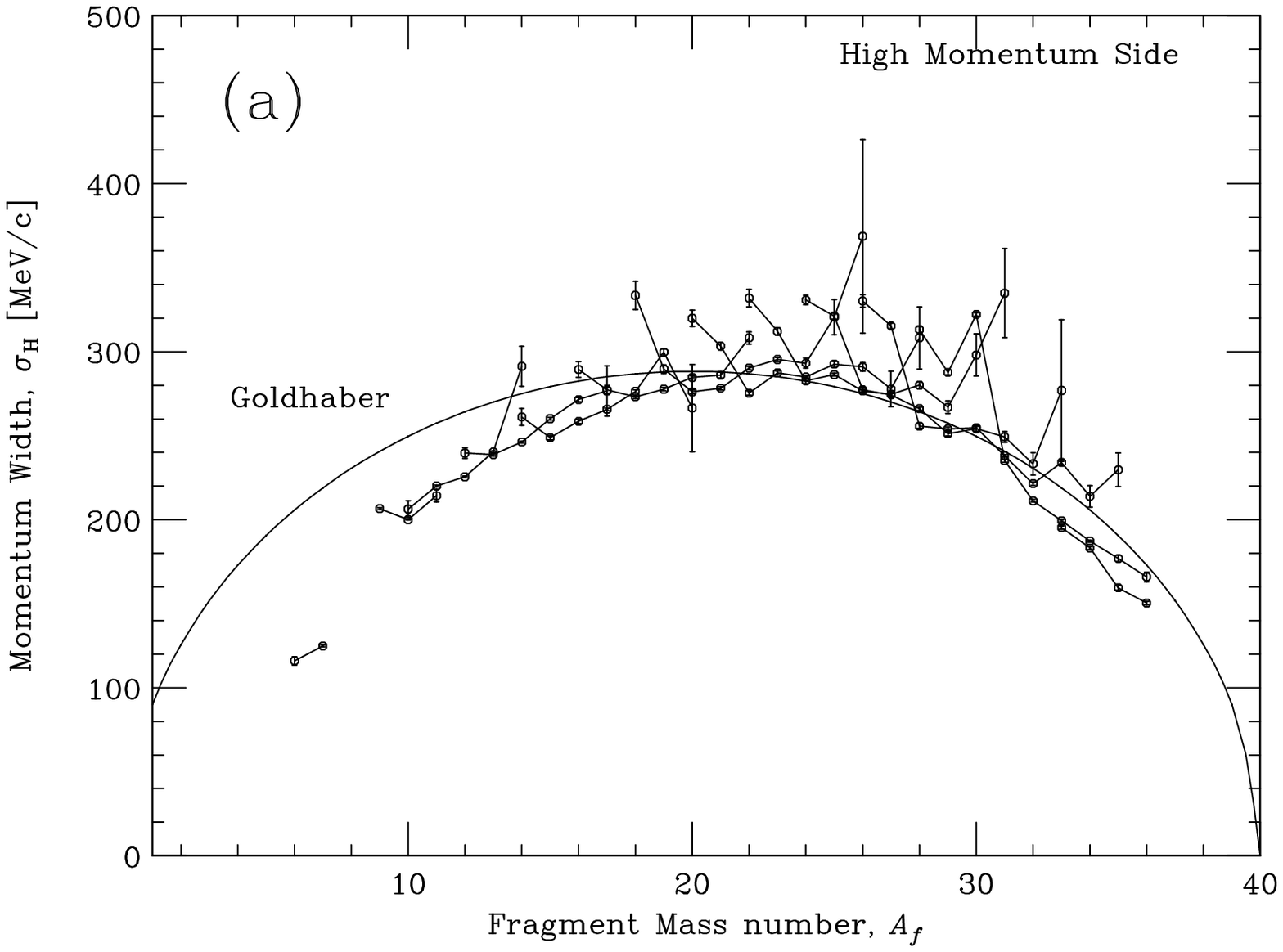}} 
~\vspace{7pt}\\
\scalebox{.40}{\includegraphics{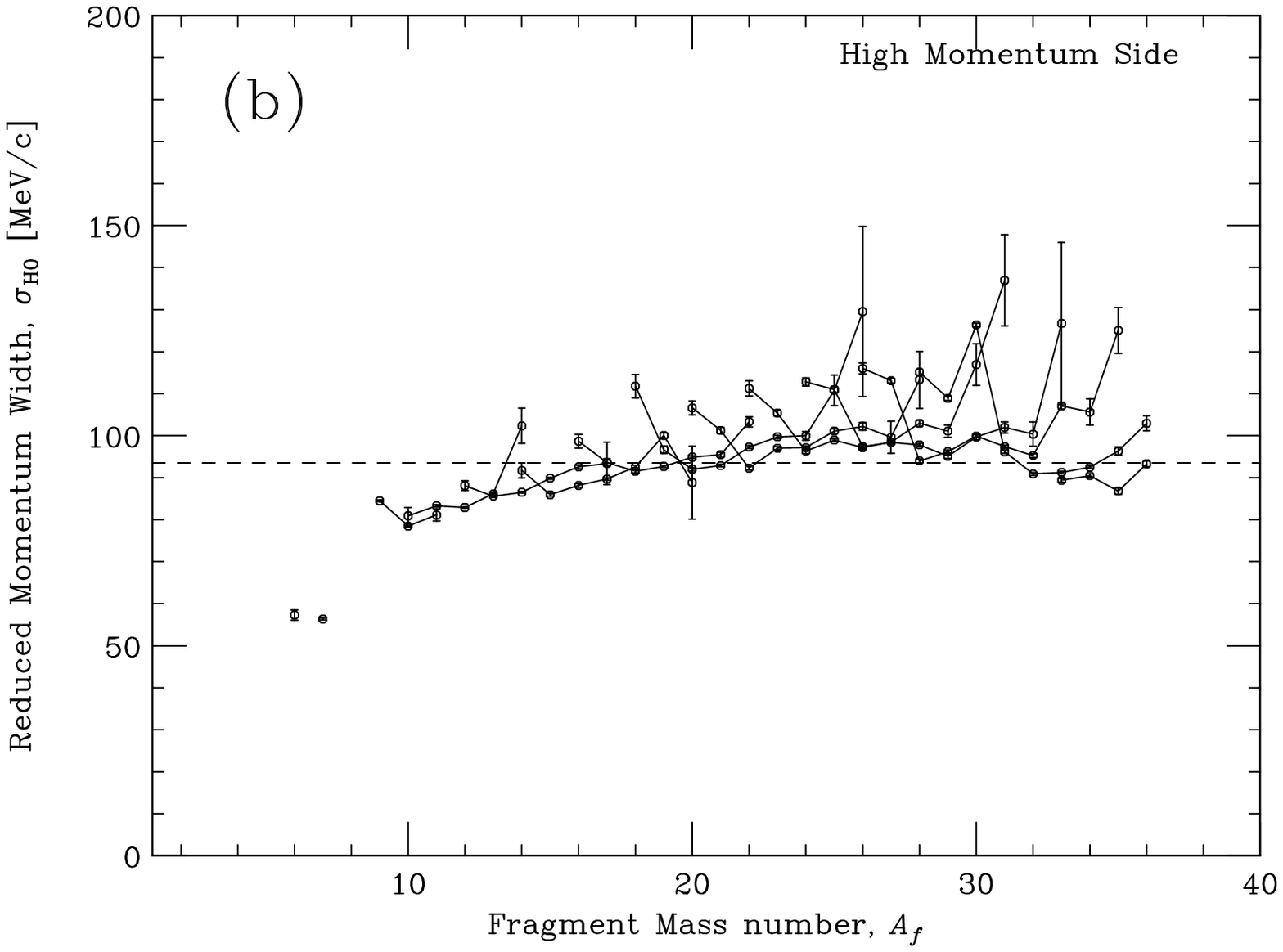}} 
\caption{\label{fig:width}~Momentum widths at high momentum side (Be target);
(a)~for ${\sigma}_{H}$ and (b)~for reduced widths
according to the Goldhaber model.}
\end{figure}
\begin{figure}[t]
\scalebox{.40}{\includegraphics{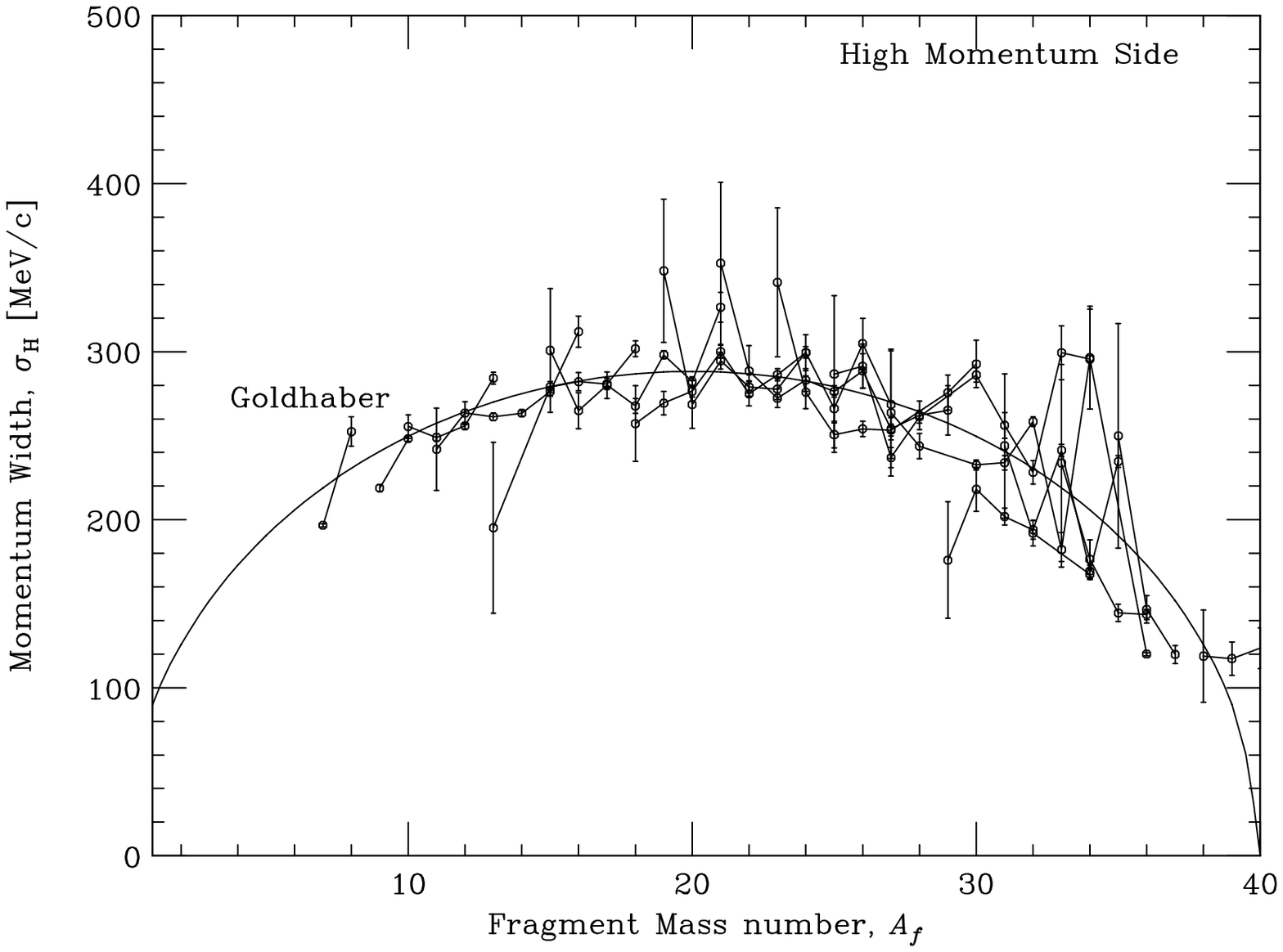}} 
~\vspace{7pt}\\
\scalebox{.40}{\includegraphics{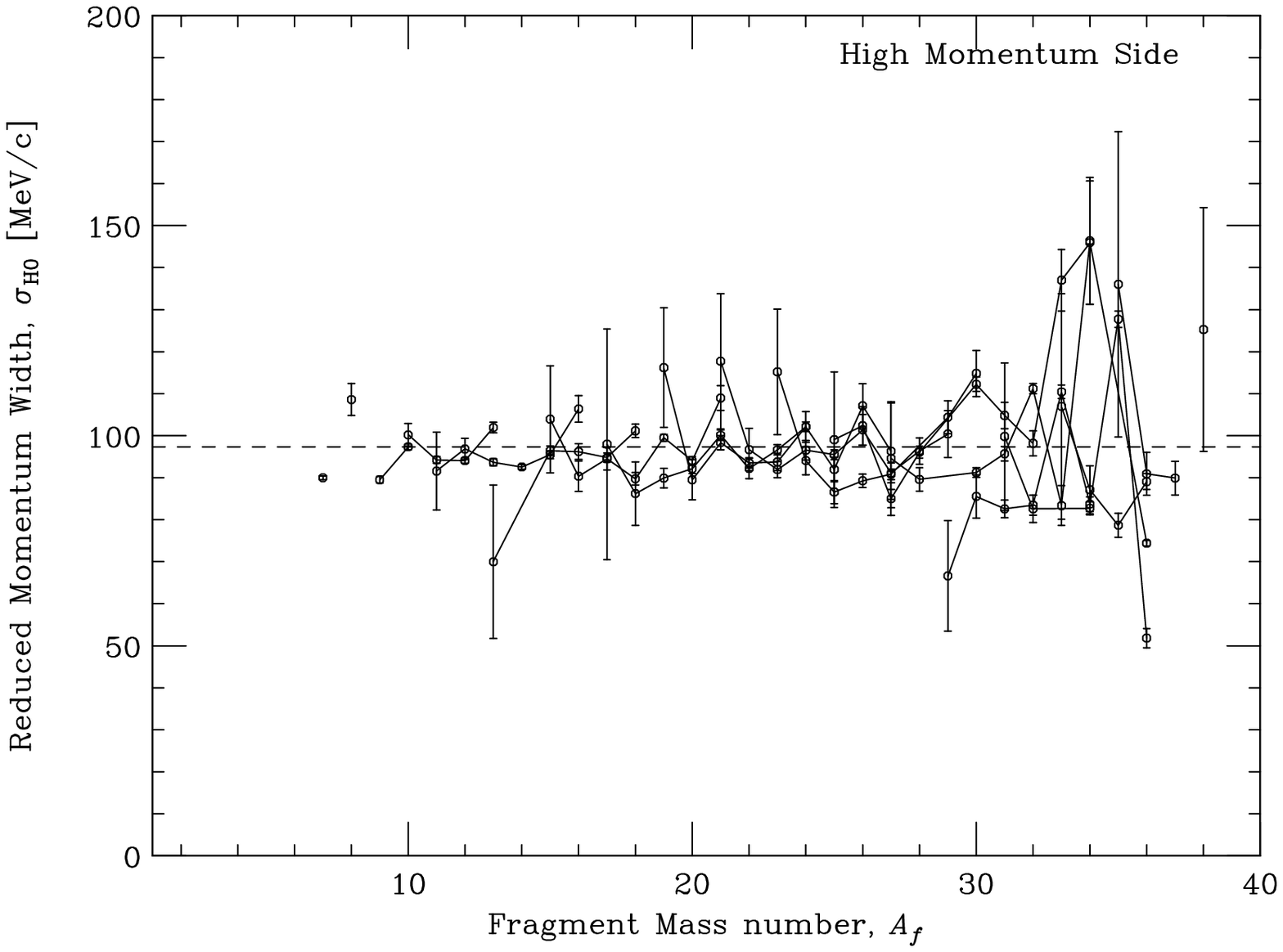}} 
\caption{\label{fig:width-high-Ta}~Momentum widths at high momentum side (Ta target);
(a)~for ${\sigma}_{H}$ and (b)~for reduced widths
according to the Goldhaber model.}
\end{figure}

The high momentum side widths ${\sigma}_{H}$ as a function of
fragment mass are shown
in Fig.~\ref{fig:width}\/(a) and Fig.~\ref{fig:width-high-Ta}\/(a),
for the Be and Ta targets, respectively.
The high momentum side widths of all the observed fragments are
compared to the formulation by Goldhaber
as follows,
\begin{eqnarray}
{\sigma}_{||} & = & {\sigma}_{0} \sqrt{\frac{{A}_{f}({A}_{p}-{A}_{f})}{{A}_{p}-1}} \label{eq:Goldhaber}
\end{eqnarray}
\noindent
where ${A}_{p,f}$ is the mass number of the projectile and fragment,
respectively. 
The solid curves are drawn with the reduced width ${\sigma}_{0}$\/=\/90 MeV/c
from experimental results at relativistic energies~\cite{Gold74}.
The deduced values of ${\sigma}_{0}$ for the fragments
in the mass range 9 to 36 are shown
in Fig.~\ref{fig:width}\/(b) and Fig.~\ref{fig:width-high-Ta}\/(b).
The dashed lines denote the mean value of ${\sigma}_{0}$
=\/93.5$\pm$${2.6}_{stat}$$\pm$${7.5}_{sys}$ for the Be target, and
97.4$\pm$${1.8}_{stat}$$\pm$${7.8}_{sys}$~MeV/c for the Ta target,
respectively.
No significant difference between both targets
was observed in the high momentum side widths.
The light fragments ${A}_{f}$$<$13 produced by using the Be target
show slightly a deviation from the Goldhaber model.
The deviation is originating from the LE-component and
the fitting function to be used.
To avoid a digression from the main purpose,
we do not follow up the deviation further.
These results are in good agreement with
high-energy experiments~\cite{Gold74}.

At relativistic energies, the reduced width ${\sigma}_{0}$ is
independent of the primary beam energy.
At lower energies,
the reduction of ${\sigma}_{0}$ has been observed~\cite{Guer85}.
The reduction mechanism has been argued by several theoretical works,
where the reason is for example due to the Pauli blocking.
Due to the effect, the ${\sigma}_{0}$ has the energy dependence
at 30-40$A$ MeV~\cite{Bor83,Mer86}, and becomes constant
up to 90$A$ MeV.
The fact that the measured ${\sigma}_{0}$ is the same as
high-energy one is consistent with this picture.

\subsection{\label{sec:l_width}Low momentum side width}
\begin{figure}[t]
\scalebox{.45}{\includegraphics{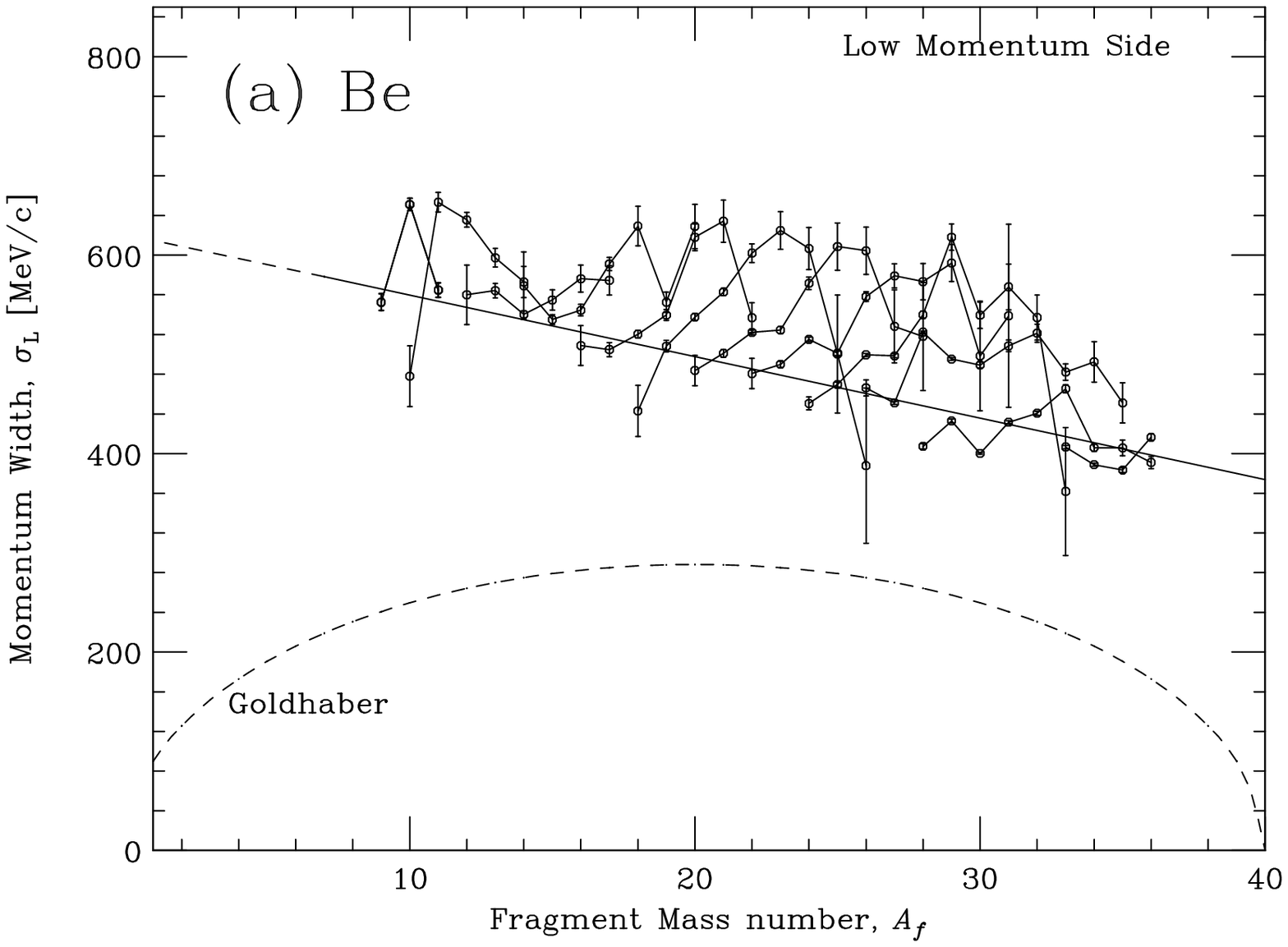}} 
~\vspace{7pt}\\
\scalebox{.45}{\includegraphics{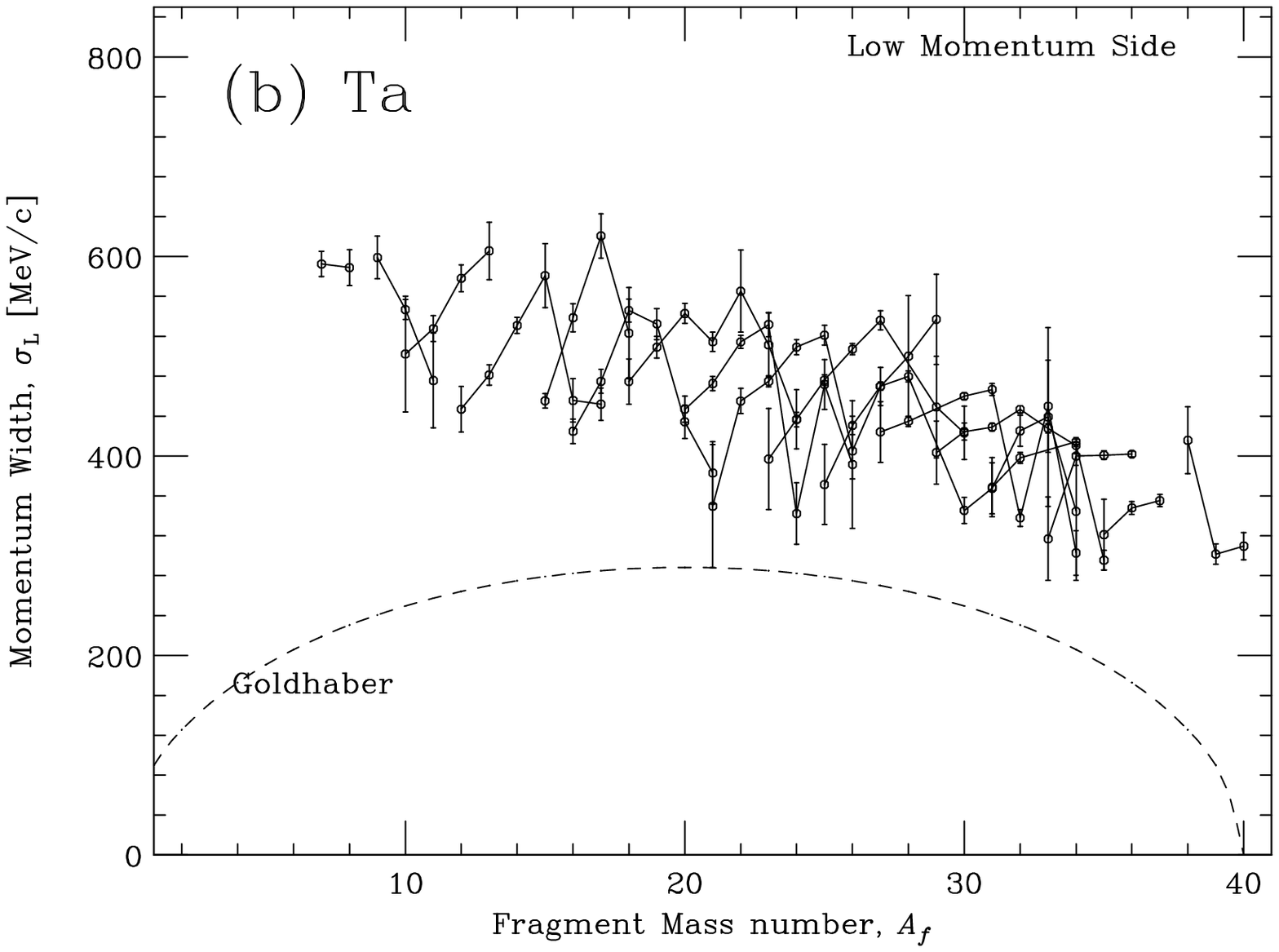}} 
\caption{\label{fig:width-low0}~Momentum widths
at low momentum side by using
(a)~the Be target and (b)~the Ta target.
The solid lines in both figures are the linear fitting of
the Ta target data.
The experimental results~(${\sigma}_{L}$) and the fitting lines
are located far from a prediction of the Goldhaber model.
The isotope chains of experimental data are also drawn as solid
lines.}
\end{figure}

The results of momentum width at low momentum side are shown
in Fig.~\ref{fig:width-low0}.
The widths of low momentum side~${\sigma}_{L}$ are plotted
as a function of fragment mass with the statistical errors.
We compared the results with the high momentum side width~${\sigma}_{H}$.
Instead of showing individual data of ${\sigma}_{H}$,
the dashed curve calculated with the Goldhaber model is presented.
The systematic error of ${\sigma}_{L}$ is estimated to be 8\%,
which is not shown in the figure.

In the Be-target data, each isotope chain has a mountain-style
structure (solid curves).
On the other hand, the low momentum widths ${\sigma}_{L}$
of the Ta data may have no such a structure.
One can understand the deviation as the target thickness effect.
Another reason for the mountain-style structure should be noted.
The momentum distribution of light fragments has the LE-component
at the low momentum side only for the $^{9}$Be target data.
In the fitting procedure by use of the asymmetric Gaussian-like function,
the LE-component may also affect the results.
Therefore, the mountain-style structure of ${\sigma}_{L}$ for fragment mass
should be ignored as a systematic error.

As seen in Fig.~\ref{fig:width-low0},
the measured widths of ${\sigma}_{L}$ are
twice wider than ${\sigma}_{H}$ approximately.
Obviously, the large width cannot be explained by the Goldhaber model.
The ${\sigma}_{L}$ may have a linear dependence as a function
of mass loss $\Delta A={A}_{p}-{A}_{f}$, which is very different from
the parabolic feature of ${\sigma}_{H}$.
No target dependence of ${\sigma}_{L}$ between the Be and Ta targets
was found in our data.
As any models for the low momentum tail have not been established,
it is difficult to discuss the low momentum tail
only from the systematics of ${\sigma}_{L}$.
Yet we should comment that the low momentum tail may not depend on
the target.

In addition, another interesting point was found in the result.
At the limit $\Delta A$$\rightarrow$$0$,
the ${\sigma}_{L}$ may not converge on 0, but 300$\sim$400 MeV/c.
This feature is different from the Goldhaber formula.
Even if the fragmentation reaction is the dominant process
for production of fragments,
these new observations may lead to an additional reaction process
necessary for intermediate-energy reactions.

In Fig.~\ref{fig:width-low0},
a fitting line was first obtained from the Ta data.
The fitting lines are drawn in both figures. 
The experimental results~(${\sigma}_{L}$) and the fitting lines
are located far from a prediction of the Goldhaber model.

What is the origin of the large width of low momentum side?
The large width is indeed produced by
an energy-loss process in the nuclear reaction.
The energy-loss process for ${\sigma}_{L}$
may be different from the ``pure'' fragmentation process, because
the width of low momentum side
strongly depends on the beam energy.
 At relativistic energies, the momentum distribution becomes symmetric
with ${\sigma}_{L}$ = ${\sigma}_{H}$.
For a low energy beam at 30$A$ MeV,
the large tail of low momentum side appears
obviously~\cite{Rus95}.
Macroscopic friction process of nuclear dissipation
is not the satisfactory mechanism for ${\sigma}_{L}$
because the momentum peak ${P}_{0}$,
which has no significant energy dependence,
may also be changed at the same time.

The energy-loss process to produce the width of low momentum side
may be explained by nucleon exchange reaction
between target and projectile~\cite{Greg86}.
For instance, transfer mechanism adds one nucleon to the
projectile or fragment, the energy per nucleon~$E$ may be changed
as ${E}_{+1n} = E\cdot A/(A+1)$,
where $A$ is the mass number before transfer reaction.
Since the velocity of projectile is reduced by
the nucleon exchange reaction,
the nucleon transfer may contribute the low momentum tail.
When the projectile gives a nucleon to the target nuclei,
the velocity of the remained nucleons in the projectile does not change
significantly.
Like the effect of momentum peak shift, the change of potential
energy in projectile affects slightly the velocity of projectile nucleus.
On the other hand, when the projectile picks up a nucleon
from the target nuclei, the projectile velocity should
be deduced
because nucleons in target nuclei are much slow on average
at the laboratory system.
Thus, we suppose the nucleon exchange reaction as an origin of
the low momentum tail.

Assumed that the probability of nucleon exchange is described as
the Poisson distribution and the average of number of
transfer nucleons in a reaction is quite small,
the transfer process does not contribute the peak shift of
fragment-momentum distribution
but the large width of low momentum tail.
 If the probability of transfer process is small on average,
the peak shift of fragment-momentum distribution,
which is related to the case of no nucleon transfer,
does not suffer from the transfer process to be independent of
the beam energy.
On the other hand, the low momentum width is sensitive to
the transfer probability.

We next discuss the linear dependence of ${\sigma}_{L}$
as a function of $\Delta$$A$.
First, we try to explain this dependence in terms of
the surface abraded. If the transfer probability is proportional
to area of the surface, the probability may be described as
a symmetric function with respect to the half of projectile
mass. However, the observed behavior has a linear dependence.
So, the abraded surface may not be directly related to
the transfer process.

Secondly, we think about the overlap volume of projectile and
target nuclei, i.e., total number of nucleons in the
participant region.
The region has almost linear dependence
as a function of ${A}_{f}$.
Thus, the linear dependence of ${\sigma}_{L}$ may be
related to
the volume of the overlap region.

\subsection{\label{sec:trans_frag}Transfer-like fragmentation}
The transfer-like fragments of $^{36}$Al,$^{37,38}$Si,$^{38,39}$P, and
$^{37-40}$K were observed at this experiment.
These fragments have more neutrons or protons than the projectile nucleus.
The fragments cannot be produced with projectile fragmentation reactions.
Neutron and proton pick-up processes are necessary for production
of the fragments.

\begin{figure}[t]
\scalebox{.40}{\includegraphics{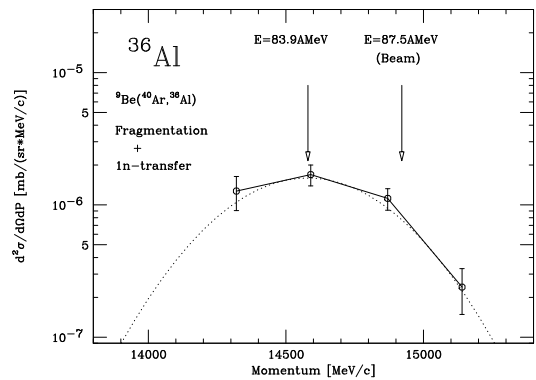}} 
\caption{\label{fig:TLF1336C4}~Momentum distribution for $^{36}$Al.
This isotope cannot be produced with only the nuclear fragmentation
process. One neutron should be picked up from target.
The measured peak momentum corresponds to 83.9$A$ MeV.
The measured momentum peak shows a larger shift
toward low momentum side, compared with that of the
same mass number, $\sim$0.9$A$ MeV.}
\end{figure}

Figure~\ref{fig:TLF1336C4} shows the momentum distribution of $^{36}$Al
acquired by using the Be target.
This isotope production needs at least one neutron pick-up process
from target nucleus.
We made an attempt to fit the momentum distribution
with the asymmetric Gaussian function.
Due to the lack of data at low-momentum tail,
we assumed that the fitting parameters of momentum widths were fixed.
The ${\sigma}_{H}$ was given from the Goldhaber formula with
${\sigma}_{0}$=90 MeV/c.
The ${\sigma}_{L}$ was assumed to be ($400\pm{60}_{sys}$) MeV/c.
This value was obtained from the systematics
as shown in Fig.~\ref{fig:width-low0}.

The fitting result is shown as a dotted curve
in Fig.~\ref{fig:TLF1336C4}.
The peak of the momentum distribution corresponds to 83.9$A$ MeV.
As the primary beam energy was measured as 87.5$A$ MeV,
the momentum peak shift was obtained as (3.6$\pm$1.2) $A$~MeV.

The measured momentum peak shows clearly a larger shift
toward low momentum side,
compared with that of the same mass number, $\sim$0.9$A$ MeV.
The large momentum shift of transfer-like fragments
was also observed in the fragments of $^{37-40}$K
requiring to proton pick-up process
in production.
%

\subsection{\label{sec:fact_check}Target dependence of cross sections}

The target dependence of cross sections was investigated
with the results of our experiments.
We have obtained the cross sections in wide range of fragment charge
for each fragment mass with small statistical and systematic errors,
and for the same projectile~($^{40}$Ar)
with two sets of targets~($^{9}$Be and $^{181}$Ta).
So we can investigate the validity of factorization
for fragmentation reactions at intermediate energies.

\begin{figure}[t]
\scalebox{.45}{\includegraphics{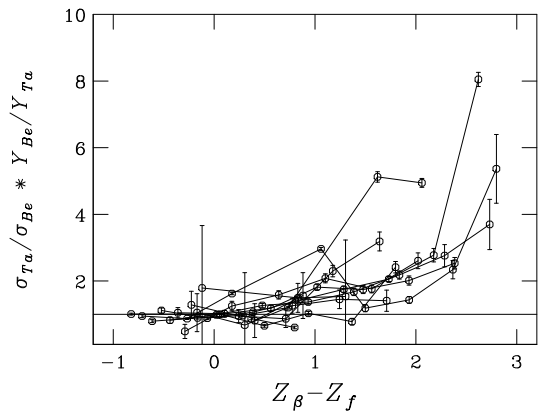}} 
\caption{\label{fig:csratio}~Ratio of the production cross sections
for each fragment
produced with Be and Ta targets. The solid lines are drawn for
the same mass number of nuclei, i.e. isobars.
The ${Z}_{\beta}$ is the $\beta$-stable charge for each isobar.
The cross-section ratio of ${\sigma}_{Ta}$(A,Z)/${\sigma}_{Be}$(A,Z)
is normalized with the mass-yield ratio
of ${Y}_{Ta}$(A)/${Y}_{Be}$(A) observed.}
\end{figure}

Figure~\ref{fig:csratio} shows the ratios of cross sections
for a fragment in Ar+Ta reactions to those in Ar+Be reactions.
The cross sections are normalized with the experimental mass yield
$Y$(${A}_{f}$) to eliminate the target size effect.
The mass yields were obtained from the sum of fragment cross sections
with the same mass number.
The ratio is shown as a function of charge difference
between the most stable charge~${Z}_{\beta}$(${A}_{f}$)
and the fragment charge ${Z}_{f}$.
Thus, the target dependence of projectile fragmentation cross sections
except the target size effect is represented
for the wide range of isotopes near and far from the stability line.
If the factorization is valid for production of an isotope,
the ratio has no ${Z}_{\beta}-{Z}_{f}$ dependence.
%

In Fig.~\ref{fig:csratio},
the ratios near the $\beta$-stability line are constant.
It corresponds
that the factorization is valid for the production of the nuclei.
According to the previous work~\cite{Olson83},
the factorization is valid for the isotopes
in ${Z}_{\beta}-{Z}_{f}$$\leq$2.
On the other hand,
the ratios increases when ${Z}_{\beta}-{Z}_{f}$ is increased.
This deviation shows that
the factorization hypothesis is clearly broken down
for neutron-rich nuclei with (${Z}_{\beta}-{Z}_{f}$)$\geq$2.

\subsection{\label{sec:epax}EPAX formula}

The EPAX was developed by using data of the spallation reactions
and heavy-ion induced fragmentation reactions
at several~$A$~GeV~\cite{Sum90}.
The cross section of a fragment with mass $A$ and proton number $Z$
produced by projectile fragmentation from
a beam (${A}_{p}$, ${Z}_{p}$) impinging on
a target (${A}_{t}$, ${Z}_{t}$) can be written as,
\begin{eqnarray}
\sigma(A, Z) & = & Y(A) \cdot W(A,Z)\\
W(A,Z)       & = & n \cdot \exp (-R \cdot {|{Z}_{prob} - Z|}^{U})
\label{eq:ChargeDist}
\end{eqnarray}
\noindent
where $Y$($A$) represents the mass yield which is the sum of the isobaric
cross sections for fragments with mass number $A$,
and $W$($A$,$Z$) describes the charge distribution which means
the cross-section distribution of a given fragment mass
which has a maximum peak at ${Z}_{prob}$.

The charge dispersion $W(A,Z)$ is described as 
$R$, ${Z}_{prob}$ and $U$ parameter.
The most probable charge, ${Z}_{prob}$, is written as,
\begin{eqnarray}
{Z}_{prob} & = & {Z}_{\beta} + \Delta + {\Delta}_{m},
\label{eq:zprob}
\end{eqnarray}
\noindent
where ${Z}_{\beta}$ is the $\beta$-stable charge for a fragment
of mass number, $A$~\cite{Marm71}, ${\Delta}$ is a proton-excess
between the stability line and the most probable line
of fragmentation reaction,
and ${\Delta}_{m}$ is the so-called ``memory effect'',
i.e., the influence of the projectile
$N$/$Z$ ratio on the fragment $N$/$Z$ ratio.
The $R$ is a function of fragment mass number, $A$, which shows
a fragment mass dependence of a steepness
controlling cross sections of isotopes from the stability line to
the drip lines.
The mass dependence has been confirmed from many combinations of
projectiles and targets.
The $U$ parameter is given as a constant for all neutron-rich isotopes so far.

Using new data obtained mainly at the GSI/FRS facility,
the EPAX was recently modified~\cite{Sum00}
slightly to tune the mass yield and the U parameter for proton-rich fragments
in the vicinity of projectiles.
The present paper focus on the neutron-rich fragments in a wide range.
Thus, this work gives the analysis and discussion using the EPAX formula
based on the original EPAX that has relatively simple functions.

The EPAX formula is valid for the ``limiting fragmentation''
regime,
where the fragmentation process is no longer energy dependent.
The energy dependence of fragmentation cross sections has been
investigated by Silberberg and Tsao~\cite{Silv73,Silv85}.
We can see the similar energy dependence in the total reaction
cross section.
The total reaction cross section has been studied both theoretically
and experimentally for more than 50 years. A detailed list of
reference is found in Ref.~\cite{Kox87}.
Compare the cross section of $^{12}$C+$^{12}$C at 90$A$ MeV with 900$A$ MeV,
the difference of cross sections is about 30\%~\cite{Bu84}.
This value is nearly equal to
the systematic error of measured cross sections in this work.
Thus, we can assume that the limiting fragmentation hypothesis
is valid in this work.

The EPAX formula follows the factorization hypothesis.
The mass yield has a target dependence, however it is limited
to the target nuclear-size effect.
The charge dispersion $W(A,Z)$ is independent of the target nucleus.

In this work, however,
the BOF has been found in the production of very neutron-rich nuclei.
An investigation of the charge distribution
is necessary to find appropriate description for
cross sections from our data.
In the charge distribution of EPAX,
the $U$ parameter is a constant of 1.65 for neutron-rich side,
and has no target dependence.
The value of $U$ parameter is very sensitive to the production
cross sections of the isotopes far from the $\beta$-stability line.
In the following, 
we reduce the $U$ parameter for each fragment mass with both targets
from our data. At the same time,
we investigate the target dependence of ${Z}_{prob}$
since ${Z}_{prob}$ is also sensitive
to the cross sections of isotopes far from the $\beta$-stability line.

\subsubsection{\label{sec:cd}Charge distribution}

A fitting of the production cross sections
was performed using the EPAX function.
The function is represented as the product of $Y(A)$ and $W(A,Z)$.
The charge distribution $W(A,Z)$ is characterized
with the most probable charge ${Z}_{prob}$, the slope constant $U$
and the width parameter $R$.
The $R$ parameter has slightly fragment-mass dependence
and an effect to the slope of charge distribution.
We have two kinds of $U$ parameters of ${U}_{p}$ for proton-rich side
and ${U}_{n}$ for neutron-rich side.
In the fitting procedure,
the values of the ${U}_{p}$ and the $R$ parameters are fixed,
originally given by the EPAX formula.
The fitting procedure was performed with the experimental data
of each fragment mass.
First, we obtain the maxima of the charge distribution
${Z}_{prob}$(EXP) as a function of fragment mass number
from the experimental data with both targets.
The values are compared with the EPAX ones, the $\beta$-stability line,
and $N$/$Z$ ratio of projectile nuclei.
Next, the $U$ parameters are deduced from the data
when we assume the same parametrization of ${Z}_{prob}$
for both of the targets.

\begin{figure}[t]
\scalebox{.45}{\includegraphics{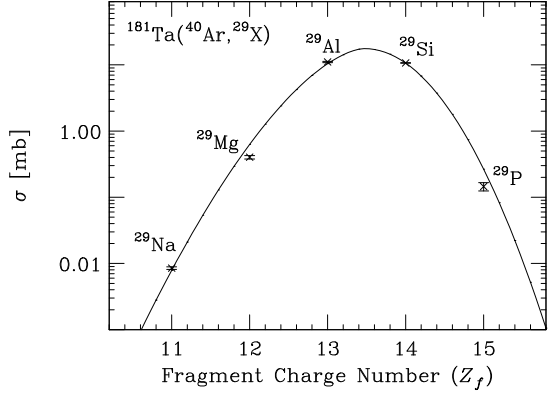}} 
\caption{\label{fig:csfit-zprob}~Charge distribution for ${A}_{f}$=29.}
\end{figure}

We have investigated the target dependence of ${Z}_{prob}$.
The measured cross sections for each fragment mass were
fitted with $Y(A)\times W(A,Z)$. The mass yield $Y$ is a constant
for a fragment mass.
Fitting parameters were $U$ parameter in neutron-rich side,
${Z}_{prob}$, and the mass yield.
Figure~\ref{fig:csfit-zprob} shows the charge distribution of
fragment mass 29, produced in Ar+Ta reaction.
The ${Z}_{prob}$ is obtained as 
13.47$\pm$0.01($\it stat$)$\pm$0.04($\it sys$).
Similarly, the ${Z}_{prob}$ value for each fragment mass
is deduced from our experimental data.
%
%

\begin{figure}[t]
\scalebox{.45}{\includegraphics{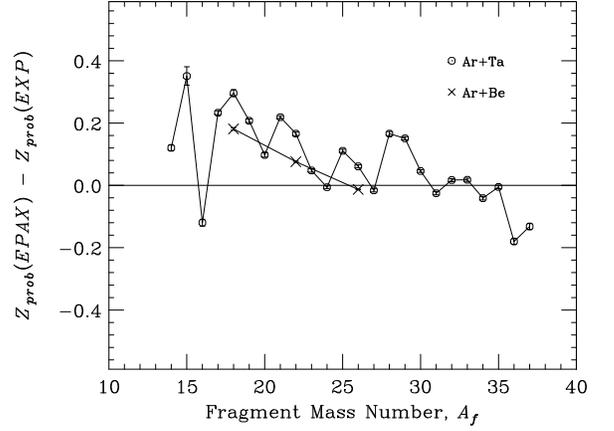}} 
\caption{\label{fig:chargedist2}~Deviation of the most probable charge from
the EPAX parametrization to experimental data
for the production targets of Be and Ta.}
\end{figure}

The experimental ${Z}_{prob}$ is compared with
the value from the EPAX formula.
Figure~\ref{fig:chargedist2} shows the deviation of
${Z}_{prob}$\/(EXP) from ${Z}_{prob}$\/(EPAX)
for Ar+Be and Ar+Ta data.
Solid lines are drawn to guide the eye.
The EPAX reproduces the ${Z}_{prob}$ very well, especially
for the mass number between 20 to 35.
The deviation of ${Z}_{prob}$ is less than 0.2.
Due to the lack of proton-rich side data,
we obtained only three values of ${Z}_{prob}$ for the Ar+Be data
in ${A}_{f}$=18, 22 and 26.
To compare both data of Be and Ta targets
shows no significant target dependence of ${Z}_{prob}$.
%
%

\begin{figure}[t]
\scalebox{.5}{\includegraphics{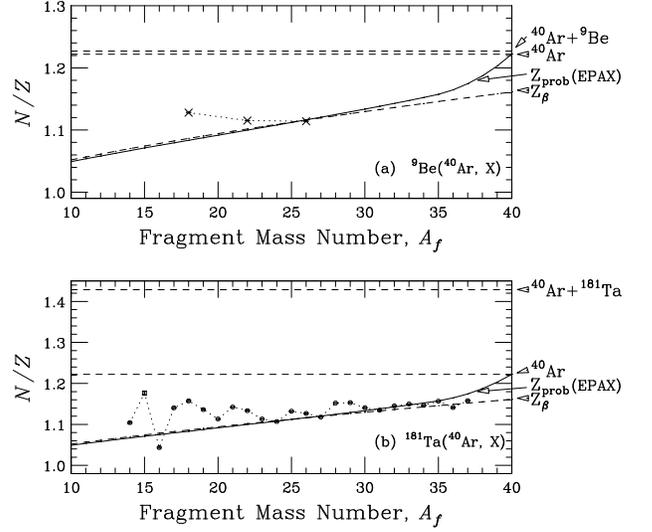}} 
\caption{\label{fig:chargedist}~Peak of charge distributions
as a function of fragment mass
for the production targets of (a)~Be and (b)~Ta.
The solid curves are the most probable charge ${Z}_{prob}$ of EPAX.
The dashed curves are the $\beta$-stable charge ${Z}_{\beta}$.
The dashed lines are the N/Z ratio of $^{40}$Ar projectile and
compounds of $^{40}$Ar+$^{9}$Be and $^{40}$Ar+$^{181}$Ta,
respectively.}
\end{figure}

Figure~\ref{fig:chargedist} shows the most probable charge
in $N$/$Z$ unit as a function of fragment mass
for the Be and Ta targets.
The most probable charge ${Z}_{prob}$ from the EPAX is
close to the $\beta$-stable charge ${Z}_{\beta}$.
The ratio ${N}_{p}/{Z}_{p}$ of the projectile is 1.22.
For the composite system of projectile and target,
the ratio is represented as $({N}_{p}+{N}_{t})/({Z}_{p}+{Z}_{t})$,
of which values are 1.23 and 1.43 for Ar+Be and Ar+Ta, respectively.
The ${Z}_{prob}$\/(EXP) is well described as
the ${Z}_{prob}$\/(EPAX) as well as ${Z}_{\beta}$.
The difference of ${Z}_{prob}$\/(EPAX) and ${Z}_{\beta}$
is mainly the memory effect ${\Delta}_{m}$.
The memory effect is not clearly seen in our data.
%
%

No significant difference of ${Z}_{prob}$\/(EXP) between
the Be and Ta targets has been observed,
however the BOF has been found in the very neutron-rich nuclei.
What does it mean?
It should be noted that the BOF had never been found
in the projectile fragmentation at other experiments.
This fact is consistent with our observation of ${Z}_{prob}$\/(EXP)
for the Be and Ta targets.
To understand the charge distribution,
we show the ${Z}_{prob}$ of projectile-like fragments
produced by the reactions at low and intermediate energies.

At low energies, the most probable charge shows
the existence of two different reaction mechanisms.
In Ref.~\cite{Gal76},
Cl isotopes were produced in 7$A$-MeV $^{40}$Ar+$^{50}$Ni
reaction and
the contour plots of the Cl isotope yield were
drawn to the fragment mass and the kinetic energy.
We clearly see two components, one corresponding to quasi elastic reactions
centered at a high energy and a mass of 39.
The second one is centered around a mass of 36 and a small kinetic energy.
The most probable charge obtained from the mass number
is near the $N/Z$ of both projectile and composite
system which corresponds to the quasi elastic and deep inelastic
reactions, respectively~\cite{Lefo78}.

An intermediate composite system has been shown
as the result of a complete damping of the relative motion
between the projectile and target nuclei.
The projectile-like fragments are produced via binary nuclear system
that collective effects dominate.
On the other hand, at high energies,
the process is dominated by individual nucleonic collisions
that described as participant-spectator models.
The values of most probable charge distinguish the reaction mechanisms
to produce the projectile-like fragments.

At intermediate energies,
D.Guerreau et al.~\cite{Guer83} reported the observation of
the systematic shift of isotope distributions
between two targets.
The isotope distributions of fragment yield were measured
in $^{40}$Ar+Ni and $^{40}$Ar+Au reactions at 44$A$ MeV.
The systematic shift of the isotope distributions for a given element of Si
in $^{40}$Ar+Au reaction was observed about 0.3 mass unit
towards the neutron rich side.
However,
their result is different from our data.
The fitting results of our charge distributions have shown
no significant difference of ${Z}_{prob}$\/(EXP) between
the Be and Ta targets.

We have already found that the factorization assumption is invalid
for production of neutron-rich nuclei.
The flexibility of the charge distribution
except the most probable charge
is the slope parameter.
Thus, we would seek for the origin of target dependence
in the slope constant $U$. Due to the BOF,
the $U$ may change from a simple constant to a complex parameter
which depends on target nuclei.

As the target dependence of ${Z}_{prob}$ has not been found,
we assume now that the ${Z}_{prob}$ can be described as
the EPAX formula for both of the targets. We make an attempt to
fit the data with the function of charge distribution.
Fitting parameters were $U$ in neutron-rich side and mass yield $Y$,
and fragment mass dependence of $U$ is tried to be obtained.
The fitting procedure
is performed for the data of each fragment-mass group, where
every 4 mass units are combined to avoid poor statistics.

\begin{figure}[t]
\scalebox{.5}{\includegraphics{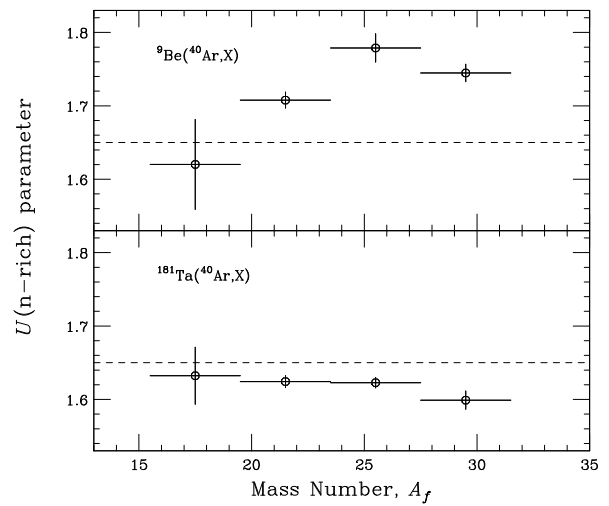}} 
\caption{\label{fig:uparam}~$U$ parameters as a function of fragment mass
for the production targets of Be and Ta.
On the assumption that the value of $U$ is same for.
every group of four sets of ${A}_{f}$ data,
the fitting results are shown with the systematic error.
The dashed lines are the values of $U$=1.65 from the EPAX.}
\end{figure}

Figure~\ref{fig:uparam} shows the $U$ parameters as a function of
fragment mass for both targets.
We plot the values with the systematic errors.
The slope parameter $U$ of EPAX is a constant of 1.65~(dashed lines).
The average of $U$ is 1.62 for the Ta target. The smaller $U$ gives
larger values of production cross sections for neutron-rich nuclei.
On the other hand, the $U$ parameter of Be target is larger than 1.65
and shows fragment mass dependence.
Our data demonstrate the BOF 
for very neutron-rich nuclei as the target dependence of
$U$ parameter.
%
%

\subsubsection{\label{sec:power}Predictive power of new parametrization}

\begin{figure}[t]
\scalebox{.5}{\includegraphics{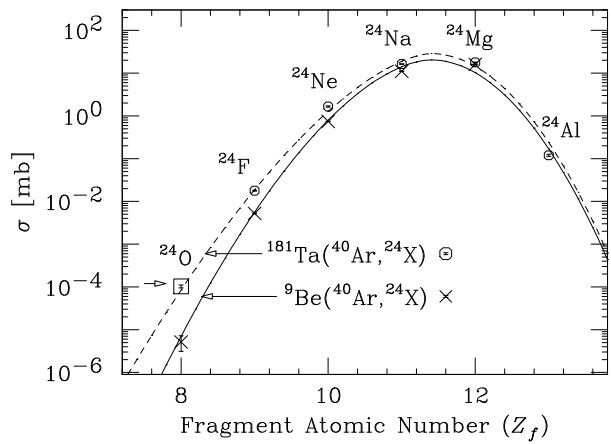}} 
\caption{\label{fig:o24}~Predictive power of cross sections of
the new parametrization.
The solid box is the data of $^{24}$O with a tantalum target.
}
\end{figure}

We have obtained the modified EPAX formula
for the nuclear fragmentation at an intermediate energy
for both Be and Ta targets.
To confirm the validity of our parametrization,
we show an example of the production cross section
predicted for extremely neutron-rich nuclei.
Figure~\ref{fig:o24} shows the predictive power of
the new parameterization. The dashed and solid curves are
the charge distribution of mass-24 isobars produced
in $^{40}$Ar+$^{9}$Be and $^{40}$Ar+$^{181}$Ta
reactions, respectively.
The solid box is the cross section of fragmentation channel
for $^{181}$Ta($^{40}$Ar,$^{24}$O).
This value acquired in another experiment~\cite{Yana03} is
in agreement with our new parametrization clearly.
%

\subsection{\label{sec:mech}Mechanism of Prefragment Production}

We have investigated the nuclear fragmentation reaction
in nucleus-nucleus collisions at intermediate energies.
The momentum distributions of projectile-like fragments
have been measured for various isotopes
including very neutron-rich nuclei.
The charge distributions of fragment cross sections
acquired with the Be and Ta targets
have revealed the BOF for very neutron-rich nuclei.
In order to explain the origin of BOF,
we discuss the mechanism of prefragment production
in terms of the AA model.
In the AA model, the production cross section of a final fragment
is determined by the charge distribution and the excitation energy
of prefragments.

First, we discuss the charge distribution of prefragments
on the basis of AA model.
The abrasion model is a macroscopic description which
gives a picture of a clean cut of the projectile nucleus
by the target nucleus~\cite{Gos77}.
The concept of geometrical separation in an overlap
or participant zone does not specify its proton-to-neutron ratio.
A few different methods to calculate the charge distribution
of the prefragments are so far proposed,
a correlation model~\cite{West79},
hypergeometrical model~\cite{Fried83},
and GDR model~\cite{Morr78}.
After all,
the charge distribution of prefragments
from these methods depends on the projectile nuclei
but has no target dependence.

\begin{figure}[t]
\scalebox{.5}{\includegraphics{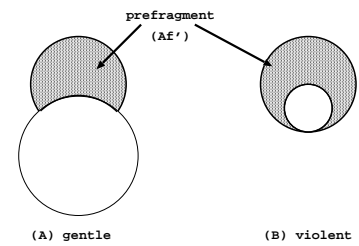}} 
\caption{\label{fig:Over2}~A schematic drawing of prefragment production.
The prefragments of the same mass number (${A}_{f}$')
are produced by projectile fragmentation reaction
with target nuclei of different nuclear radius.
For the large target nuclei,
the large impact parameter leads to (A)~gentle reaction,
and (B)~violent reaction is expected for the small impact
parameter with the small target nuclei.}
\end{figure}

Next, we discuss the excitation energy of prefragments.
In an early model~\cite{Bow73},
the average excitation energy of a prefragment after abrasion was estimated
from the additional surface energy due to the excess surface area.
Several studies have been made on the source of the excitation energy
~\cite{Huf85,Wil87,Gaim91}.
According to the researches, the surface energy is not the main source
of the excitation energy,
but, one of the energy sources for prefragment excitation.
At least, the target dependence of the excitation energy can be seen in
the surface energy.
Figure~\ref{fig:Over2} shows a schematic view of prefragment production.
The prefragments of the same mass number ${A}_{f}$' are produced
by the projectile fragmentation reaction with target nuclei
of different nuclear radius.
If a target nucleus has a large size, the impact parameter is
also large to produce the prefragment~(case A).
On the other hand, the small size of a target nucleus
caused the violent collision to form the prefragment with the same
mass number~(case B).
In the geometrical model,
we calculated the surface energy of prefragments
in Ar+Be and Ar+Ta reactions.
As the result,
large difference of ${E}_{s}$ is presented for both of the targets.
As for the prefragments of ${\Delta}A=20$, the values of ${E}_{s}$
are approximately 90~MeV and 10~MeV for the Be and Ta targets, respectively.
The target dependence of prefragment excitation energies
stems from the additional surface energies in the AA model.
%
%

We observed no significant target dependence of momentum peak shift
in the data for the projectile-like fragments of ${A}_{f}\geq$20.
Assumed that nucleons are emitted homogeneously from the hot prefragment
in the ablation process,
the velocity of a prefragment does not change on average
when it becomes the final fragment.
The results of momentum peak shift show that the energy losses
in the fragmentation process for the fragment of $\Delta A=20$
were 60$\sim$80 MeV for both of the targets.
The energy losses are not correlated to the additional surface
energies on the calculation.
Thus, we cannot deny the possibility
that the BOF originates from
the target dependence of the charge distributions for prefragments
except the excitation energies.

To apply this idea to the BOF problem,
we suggest two mechanisms of a fluctuation
giving for the charge distributions of prefragments.
One is the nucleon exchange process during the fragmentation process,
and another is the effect from the difference of nucleon-nucleon
total cross sections~(${\sigma}_{pp}$,${\sigma}_{nn}$ and ${\sigma}_{pn}$).

We observed the nucleon exchange process
during the fragmentation process in our data.
The fitting results of ${P}_{0}$ and ${\sigma}_{H}$ carry
the information at the abrasion stage
and supporting that ``pure'' projectile fragmentation is
a dominant process at intermediate energies.
As for ${\sigma}_{L}$, the larger values of ${\sigma}_{L}$
than ${\sigma}_{H}$ may be related to the energy loss due to the
transfer process.
Unlike the behavior of ${\sigma}_{H}$,~${\sigma}_{L}$ has
a linear dependence as a function of fragment mass.
The linear dependence may be connected from both
projectile and target nuclei.
So, the source of transferred nucleons can be assumed to be
the participant region.
Since the Ta target nucleus is more neutron-rich
than the Be, the probability of neutron transfer
should be large.
The nucleon exchange reaction between projectile and target nuclei
during the abrasion process
may break the factorization of fragment production cross sections.

The difference of the total cross sections
between proton-to-proton~(${\sigma}_{pp}$)
and proton-to-neutron~(${\sigma}_{pn}$)
may bring the fluctuation of prefragment charge distributions.
As a fact known well,
the ${\sigma}_{np}$ is larger than
the ${\sigma}_{nn}$ and ${\sigma}_{pp}$
when the nucleon energy is less than 500 MeV.
In projectile fragmentation process,
a neutron-rich target nucleus easily knocks out protons
in the projectile nucleus,
so that the production of neutron-rich prefragments is promoted.

The discussions so far are qualitative.
However, they gives several predictions.
First, if the nucleon exchange process is just the reason of BOF,
the probability of nucleon exchange reaction becomes large
at low energies. The nuclear fragmentation experiments
with lower incident beam energies than at our experiment
may reveal the large BOF effect for very neutron-rich nuclei.
Secondly, if the difference of N-N reaction cross sections
causes the BOF,
the nuclear fragmentation experiment with 100-800$A$ MeV beams
should be carried out to confirm the BOF
for very neutron-rich nuclei.
Since the cross section curves of ${\sigma}_{pp}$ and ${\sigma}_{pn}$
cross at 500$A$ MeV,
the difference of U-parameter between the Be and Ta targets
might change to be reversed.
The further investigations of fragmentation cross sections
at around 500$A$ MeV are of great interest.

\section{\label{sec:sum}SUMMARY AND CONCLUSION}

The projectile fragmentation reactions at intermediate energies
have been investigated
using a 90-94$A$ MeV $^{40}$Ar beam
at RIKEN-RIPS.
We paid our attention to the target dependence.
Measurement of longitudinal momentum distributions
of projectile-like fragments
within a wide range for fragment mass and its charge
including very neutron-rich nuclei
has been performed
with $^{9}$Be and $^{181}$Ta targets.
From the momentum distribution of fragments,
a parabolic mass dependence of momentum peak shift
was observed in the data of both targets,
and
a phenomenon of fragment acceleration was observed only in
the Be-target data.
A linear dependence of the low momentum tail
as a function of removed nucleons
was found for both targets.
As a possible origin of the low momentum tail,
we suppose the nucleon exchange reaction.
We observed large target dependence of the cross sections
to produce very neutron-rich nuclei with (${Z}_{\beta}-{Z}_{f}$)$\geq$2.
The deviation shows that the production cross sections
of very neutron-rich nuclei far from the stability line
do not factorize.
The reaction mechanism was discussed, however the discussions so far
are qualitative.
It is much interesting to investigate the target dependence of
the cross sections at higher energies.
~\\

\begin{acknowledgments}
The authors wish to thank the RIKEN Ring Cyclotron staff and crew
for their cooperation.
We are also indebted to Dr. A.G. Artukh, Dr. G.A. Souliotis,
and Prof. T. Shimoda for stimulating discussions.
One of the authors (M.N.) acknowledges the finantial support
of the RIKEN Special Researchers' Basic Science Program.\\
\end{acknowledgments}
~\\
~\\
~\\



\end{document}